\documentclass[journal=jacsat,manuscript=article]{achemso}

\usepackage[version=3]{mhchem} 
\usepackage{xcolor} 
\usepackage{amsmath}
\usepackage[utf8]{inputenc}
\usepackage{relsize}

\author{Joshua A. Burrow}
\affiliation{Department of Electro-Optics, University of Dayton, Dayton, OH, USA}
\alsoaffiliation{Department of Electrical Engineering, Brown University, Providence, RI, USA}
\email{joshua_burrow@brown.edu}

\author{Md Shah Alam}
\affiliation{Department of Electro-Optics, University of Dayton, Dayton, OH, USA}

\author{Evan M. Smith}
\affiliation{KBR, Inc., Beavercreek, OH 45431, USA}

\author{Riad Yahiaoui}
\affiliation[UIC]{Department of Electrical \& Computer Engineering, University of Illinois Chicago, Chicago, IL, USA}

\author{Ryan Laing}
\affiliation{Department of Electro-Optics, University of Dayton, Dayton, OH, USA}

\author{Piyush J. Shah}
\affiliation{Apex Microdevices LLC,
West Chester, OH 45069, USA}

\author{Thomas A. Searles}
\affiliation[UIC]{Department of Electrical \& Computer Engineering, University of Illinois Chicago, Chicago, IL, USA}
\altaffiliation{Department of Physics, Massachusetts Institute of Technology, Cambridge, MA, USA}

\author{Shivashankar Vangala}
\affiliation{Air Force Research Laboratory, Sensors Directorate, Wright-Patterson AFB, Ohio 45433, USA}

\author{Joshua R. Hendrickson}
\affiliation{Air Force Research Laboratory, Sensors Directorate, Wright-Patterson AFB, Ohio 45433, USA}

\author{Andrew Sarangan}
\affiliation{Department of Electro-Optics, University of Dayton, Dayton, OH, USA}
\author{Imad Agha}
\affiliation{Department of Physics, University of Dayton, Dayton, OH, USA}
\alsoaffiliation{Department of Electro-Optics, University of Dayton, Dayton, OH, USA}
\email{iagha1@udayton.edu}
\title[An \textsf{achemso} demo]
  {Chiral Phase Change Nanomaterials}

\abbreviations{IR,NMR,UV}

\begin{document}

\begin{abstract}

Chiral nanostructures offer the ability to respond to the vector nature of a light beam at the nanoscale. While naturally chiral materials offer a path towards scalability, engineered structures offer a path to wavelength tunability through geometric manipulation. Neither approach, however, allows for temporal control of chirality. Therefore, in the best of all worlds, it is crucial to realize chiral materials that possess the quality of scalability, tailored wavelength response, and dynamic control at high speeds. Here, a new class of intrinsically chiral phase change nanomaterials (PCNMs) is proposed and explored, based on a scalable bottom-up fabrication technique with a high degree of control in three dimensions. Angular resolved Mueller Matrix and spectroscopic ellipsometry are performed to characterize the optical birefringence and dichroism, and a numerical model is provided to explain the origin of optical activity. This work achieves the critical goal of demonstrating high-speed dynamic switching of chirality over 50,000 cycles via the underlying PCNM.

\end{abstract}

\section{Introduction}
Nature's ability to mass-produce chiral elements has inspired engineers and scientists to adopt the design concepts for preferential light-matter interactions and experimental characterization techniques by exploiting the interaction between electromagnetic radiation and sub-wavelength natural or synthetic chiral structures. Mathematically speaking, a chiral object cannot be mapped to its image by rotations and translations alone; it requires a reflection transformation across a mirror plane. Unfortunately, for most natural or synthetic media, the optical response is typically fixed after fabrication making it highly challenging to dynamically tune the chirality\cite{Qui2018IntrinsicChiralityReview}, which would open the door towards an entire suite of fundamental and applied interdisciplinary studies\cite{Mun2020}. Recently, considerable effort has been expended to control handedness in nano chiral materials by exploring plasmonics \cite{StereoMeta20_AM,Zhao2012,plamonicsCD,plasmonicAOM} and high index dielectrics \cite{GradientAFM08,Zhu2018} where the refractive index is fixed, resulting in a need for mechanical deformation and reconfigured particles for tunable opto-chirality\cite{TunableDielectricsAlu21}. 

Alternatively, phase change materials, such as Ge$_2$Sb$_2$Te$_5$ (GST) possess a controllable refractive index. The tunable glass chalcogen can be reversibly cycled between amorphous (aGST) and crystalline (cGST) phase states upon thermal actuation in nanoseconds or less. 
With each state having unique and vastly different optical and electrical properties, GST has served as an inexpensive, robust, and scalable material for tunable electronic and photonic platforms. As an example, GST has gained tremendous traction due to the vastly differing high index optical properties between amorphous and crystalline states in the near- and mid-infrared regimes. In fact, there has been a wide range of reconfigurable photonics demonstrations derived from GST over the last several years focusing on amplitude and phase modulation governed by Mie resonances \cite{RuizdeGalarreta:20,PopNanoLetts21} for beam steering \cite{GalarretaBeamSteering,Zhang2021}, color modulation \cite{RiosAM} and tunable integrated photonic memories \cite{Rios2015}.
In the context of circular polarization control, recent studies investigating circular dichroism (CD) in chalcogenide PCMs include hybrid metal-PCM structures which exhibit strong tunability in the IR regime using gammadion chiral phase-change metamaterials \cite{Cao13}, and switchable CD in periodic absorbers\cite{Ding21} and three-dimensional layered split ring resonators \cite{Yin2015ActiveChiralPlasmonics}.
These realizations demonstrate active CD control in the IR regime, which can be very useful for vibrational CD applications where fabrication constraints are less critical. 
For visible applications, Shanmugam et al., demonstrated a programmable CD response by atomically encoding handedness into isotropic doped and undoped GST bulk thin films via 15 mJ/cm$^{2}$ circularly polarized light pulses \cite{PhotoInducedGST2019,Borisenko2015}. However, the CD signatures are minimal (1.16$^\circ$ ellipticity) due to the short optical path length and low chirality parameter of the optically induced medium.


While, to date, considerable effort has been expended to demonstrate nanostructures of a particular handedness, these structures require top-down, low-yield lithographic patterning. As an alternate approach in this work, a bottom-up fabrication technique for achieving nanostructure architectures based on GST with controllable three-dimensional shapes is developed. The versatility of this approach enables the facilitation and realization, at high yields, of chiral phase change nanomaterials (PCNMs) which can bolster distinct and beneficial properties relevant for selective coupling and high-speed polarization control of exotic light beams. 
\\ 
\hspace*{5mm}
In more detail, intrinsic chiral structure design and its implementation via glancing angle evaporative deposition of GST are introduced. Figure \ref{fig:conceptandesign}a displays the schematic for the approach where LCP (blue) and RCP (red) exhibit dissimilar transmissive polarization properties. Moreover, the amorphous and crystalline states exhibit unique optochiral properties. A bright-field micrograph imaged in transmission mode from a white light source is shown in Fig. \ref{fig:conceptandesign}b where the light and dark regions correspond to aGST and cGST, respectively. The bright and dark regions correspond to small and large chirality, where the Archimedean spiral is produced by local optical switching. Mueller Matrix characterization of the phase dependent optical activity is presented for the normal and oblique angle of incidence of thermally activated chiral PCNMs. Circular birefringence and dichroism are presented and compared with numerical calculations which indicate preferential absorption strictly between circular eigenpolarization states at normal incidence.
\begin{figure*}[ht!]
\centering
\includegraphics[width=0.9\linewidth]{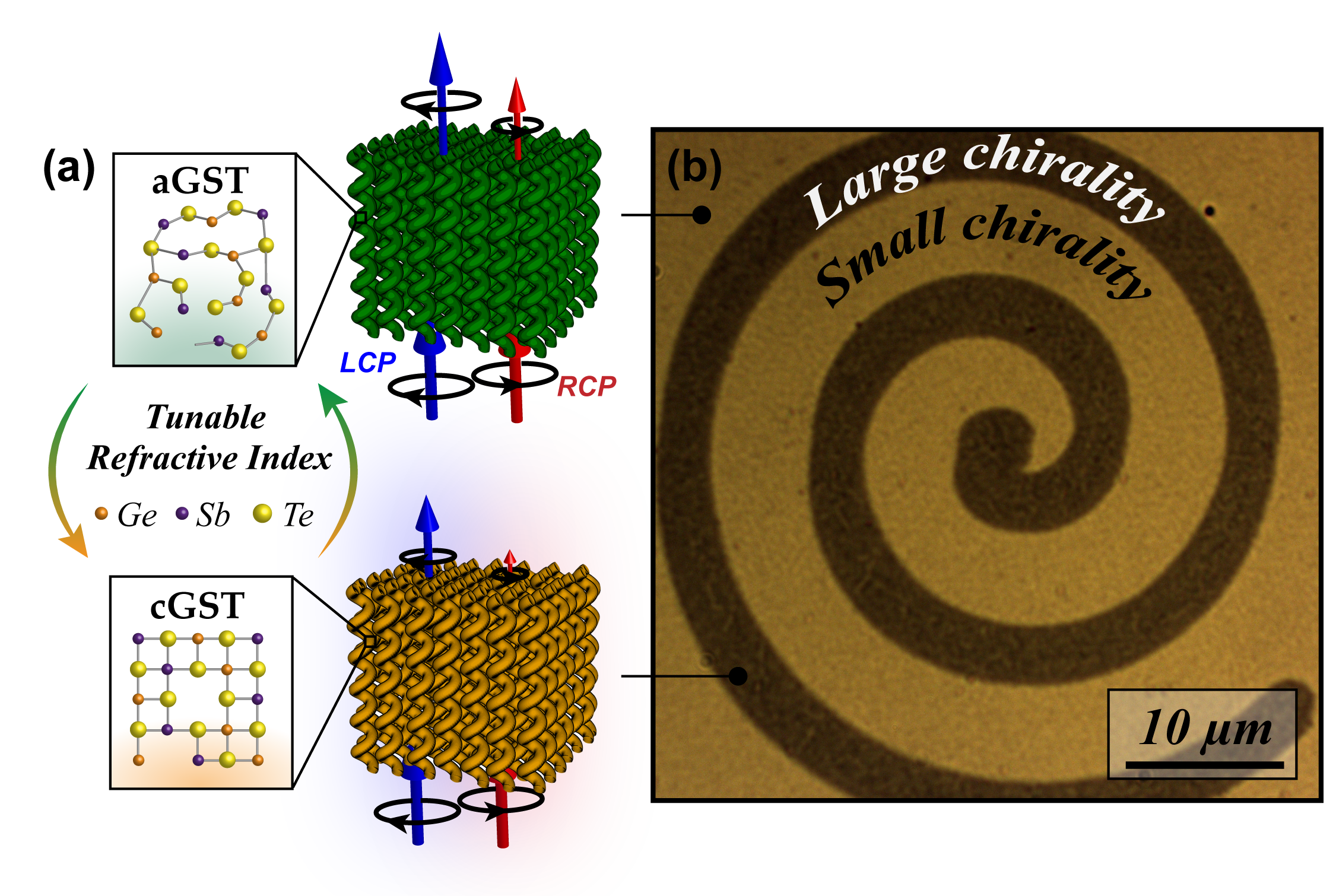}
\caption{Tunable circular dichroism (a) Schematic diagrams of the tunable chiral phase change nanomaterials for aGST (green) and cGST (gold), (b) Bright field microscope image of aGST (bright) and optically switched cGST (dark) regions.} 
\label{fig:conceptandesign}
\end{figure*}


Furthermore, a quantitative analysis is performed on the near-field coupling on and off resonance where stronger coupling is observed when the circular polarization state matches the handedness of the GST structure for a particular wavelength regime. Finally, the dimensionality of the individual helices, which matches that of state-of-the-art phase change random access memories (PCRAMs), enables the material to \textit{reversibly} switch-  point-by-point - the chiral nanostructures between their amorphous and crystalline states over $50,000+$ cycles, corresponding to reversible, high-speed amplitude and wavelength modulation of the chiral response as shown through rigorous metrology.

\section{Self-assembled GST nanomaterials enabled by shadowing mechanism}



The fabrication approach implemented here is the glancing angle deposition (GLAD) technique, which is a proven and robust method for sculpting various materials into a wide variety of exotic nanostructured columnar films \cite{hawkeye14} via electron beam evaporation. Unfortunately, there is very limited reporting on chalcogenide glass material growth via GLAD, as it presents particular challenges owing to the uncertainty of stoichiometry conservation in the evaporation of alloys.\cite{STARBOVA1997261_As2S3,Bhardwaj2007GeSe2,GSSnanorods} 
In this work, nevertheless, the  
aim is to generate cylindrical helical nanostructures for non-mechanical tunable optical activity exploration enabled by the high-speed structural phase transition in GST through a systematic study on the self-assembly growth process. Growth analysis is first performed on e-beam evaporated Ge$_2$Sb$_2$Te$_5$ onto a stationary substrate mounted at normal incidence, and verification of the stoichiometry  is enabled by energy dispersive X-ray spectroscopy (EDS) and UV/VIS ellipsometry.



\begin{figure*}[ht!]
\centering
\includegraphics[width=1.0\linewidth]{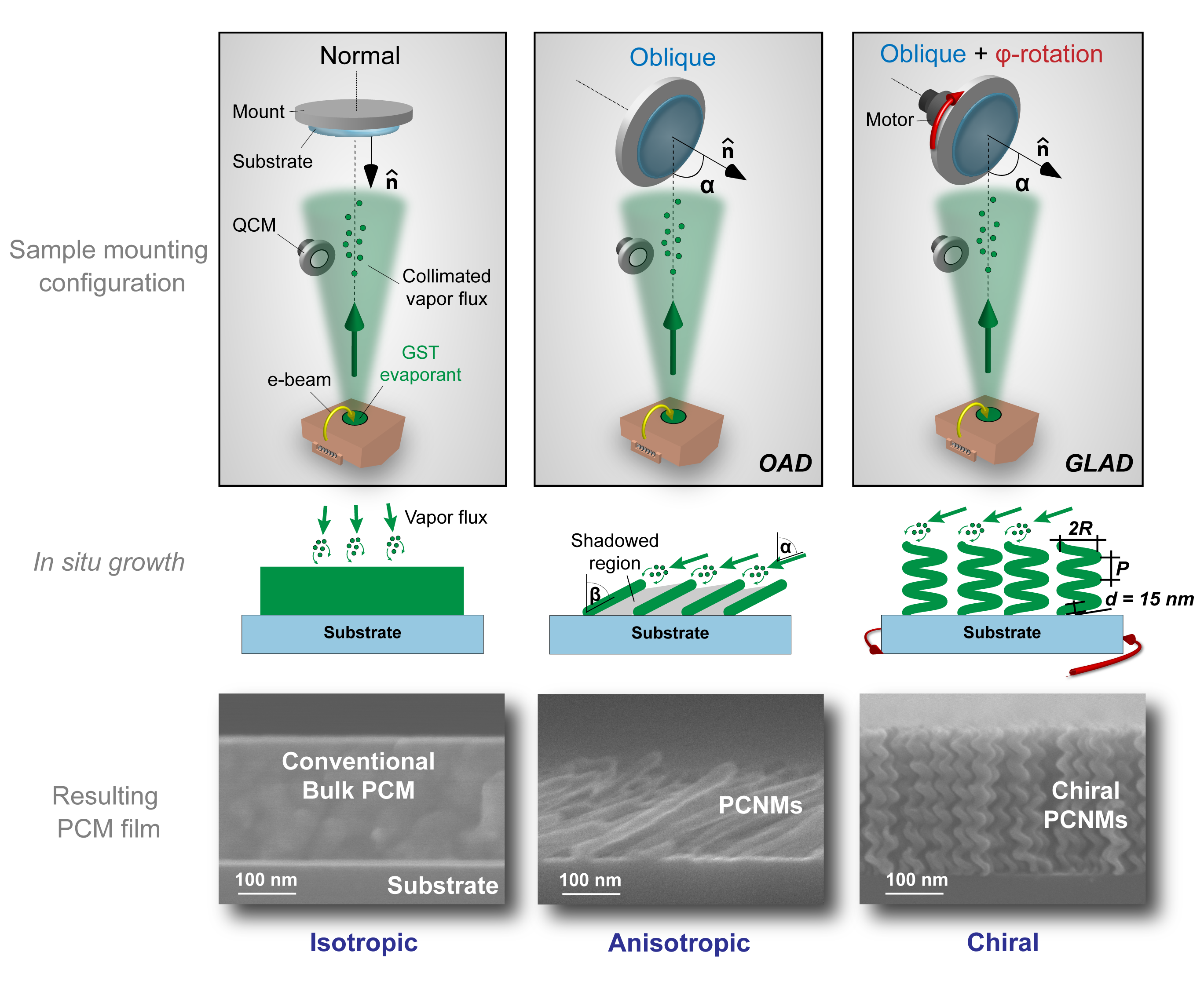}
\caption{(Top) Sample mounting configuration in electron beam evaporation chamber, (Middle) Schematic of evaporative growth growth, and (c) SEMs of resulting PCM thin films.}
\label{fig:samples}
\end{figure*}
The results of an ellipsometric gradient-index fit reveal a Ge$_3$Sb$_1$Te$_4$ (herein, referred to as GST) 
stoichiometry on average, with an increasing Te/Ge ratio as a function of distance from the substrate. 
GST is highly dependent on the deposition method and chamber conditions\cite{ZhangeAPLPerspectives2021}, thus it is worth noting that material properties exhibited by evaporated GST are very similar to our previously reported undoped sputtered GST films. \cite{GuoNidoped,GuoWdoped} Supplemental Note S1 compares the optical properties of evaporated and sputtered bulk thin films where evaporated films exhibit slightly lower refractive index and higher extinction, when compared with conventional sputtered GST.


Next, Oblique Angle Deposition (OAD) is employed to obtain directional tilted GST nanorods. Here, the sample is mounted at an oblique angle $\alpha$ with respect to the vapor flux. A series of angled depositions are carried out for discrete $\alpha$ values, including 65$^\circ$, 75$^\circ$, and 85$^\circ$ for a fixed deposition rate and shown in Supplemental Note S2. 
Next, the GLAD method is implemented, which is an extension of the OAD technique, where in GLADs the substrate is translated, rotated and/or titled during the growth processes to finely control the assembly process of the nanorods. Figure \ref{fig:samples} depicts chamber configurations that have a sample mounted at normal incidence, oblique incidence, and oblique incidence where the substrate is continuously rotated during deposition (highlighted by the vector). Prior to the GLAD, $\alpha$ is fixed to 85$^{\circ}$ and held constant throughout the programmed growth process. During the deposition the substrates is rotated at a constant angular velocity $\phi$ and the deposition rate $r$ is fixed and monitored using a current feedback mechanism via the quartz crystal microbalance (QCM).  




Initially, adatoms form a non-uniform distribution of nano clusters that act as shadows for neighboring nanosites. Owing to low mobility and limited surface diffusion, GST readily forms distinct nanostructures at room temperature.  
According to the Structure Zone Model for extreme shadowing conditions\cite{MUKHERJEE2013158}, GST shadowed growth falls into Zone II where distinct rods are formed at room temperature because GST has a homologous temperature $\theta = T_{s}/T_{m} \approx 0.34$ where T$_{s}$ and T$_{m}$ are the substrate and GST melting temperatures, respectively. Furthermore, by controlling the ratio ($\gamma$) between the sample rotation and deposition rates, one may precisely tune the helical pitch, $P$ and radius, $R$ of the revolving structures; however, the nanorod diameters $d$ are constant at 15 nm. Complete details of the resulting chiral CPNMs morphology is discussed in Supplemental Note S3.



\section{Optical-activity medium characterization}

\subsection{Normal incidence spectroscopy and FDTD modeling}
\begin{figure*}[ht!]
\centering
\includegraphics[width=1.0\linewidth]{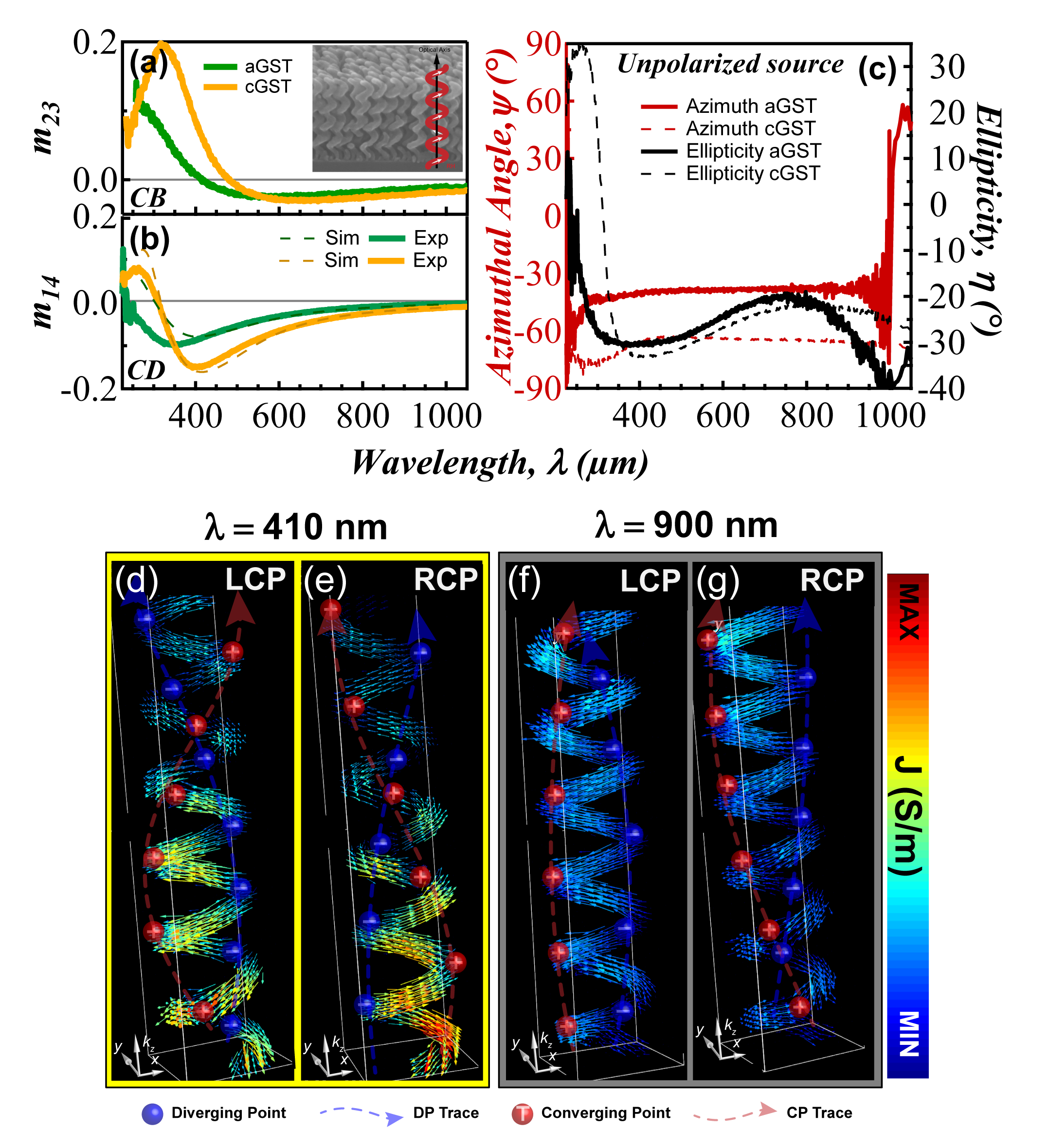}
\caption{Measured (a) m$_{23}$ and (b) m$_{14}$ of aGST (green) and cGST (gold) chiral nanorods, where dashed lines are FDTD calculations (c) azimuthal angle and ellipticity. (d)-(g) are current distribution plots of max and min CD spectral responses.}
\label{fig:CDevoluition}
\end{figure*}
\hspace*{5mm} Transmissive Mueller Matrix spectroscopy is employed to characterize the optical activity of  chiral PCNMs where the pitch $P = 2R = 45$ nm and the number of revolutions $N=5$. The measurements are performed using a dual compensator spectroscopic ellipsometer (JA Woollam RC2) with incident angle and in-plane rotation sweep capabilities. Focusing objectives allow typical spot sizes 3-4 mm that can be reduced to 100 $\mu$m in reflection mode at $55^\circ$. 

For amorphous GST, there exists a peak circular birefringence (CB) 250 nm and the largest CD amplitude is observed at 375 nm.
Upon thermal annealing the chiral molecules to the crystalline state, the measured CD spectrum is enhanced and exhibits a  redshift in the broad CD profile as shown in Figs. 3a and b. The $m_{14}$ is in good agreement with the numerical calculations where the attenuation is calculated from the normalized difference from spectra calculated from LCP and RCP excitations. The full Mueller Matrix spectra can be found in Supplemental Note S5. Additionally, the output state of polarization for a randomly polarized source is given by the azimuthal angle $\psi$ and ellipticity $\eta$ as plotted in Fig. 3c. 
Upon the phase change process, the enhancement, CD resonance shift, and broadening, can be attributed to the shift in the complex refractive index when the atomic bonds are reconfigured into a FCC crystal lattice.

\vspace{5pt}
\hspace*{5mm}The difference in CD amplitude between numerical calculations and measurements can be interpreted as a result of the aperiodic stochastic distribution of PCNMs. 
Homogeneous ordered structures result in more significant transmission and reflection coefficients due to an enhancement of electromagnetic activity, but typically the resonances are wavelength invariant. 
To account for signal attenuation, the long-range order is relaxed by assuming periodic boundary conditions and averaging the transmission coefficient over a range of square periodicities within one standard deviation of the normally distributed nanorods, which is highlighted in Supplemental Note S4. 
Below 410 nm the CD begins to decrease, then exhibiting a CD reversal point near 320 nm. 
At longer wavelengths, the CD signal exponentially decays to near-zero CD.
Next, the current density $J$ is accessed inside the chiral PCNMs when excited by each circular eigenpolarization state.
Figure 3d-g plots the displacement current density at 410 and 900 nm for LCP and RCP where the vector field plot distributions have been scaled for visual clarity and plotted on a common colorbar for ease of comparison. The current distribution is calculated as  $J = -i\omega(\epsilon - \epsilon_{0})E$ where $E$ is the electric field, $\omega$ is the angular frequency of electromagnetic radiation, $\epsilon$ is the material permittivity, and $\epsilon_0$ is the permittivity of free-space.
At maximal CD, one can observe that the circularly polarized light couples only to a helix of matching handedness; thus, a stronger $J$ is observed in Fig. 3d relative to an unmatched polarization in Fig. 3e. 

Thus, when the circular polarization matches the handedness of the helix it impinges on, large opposing loop currents are observed in that helix (aGST-LCP). If, however, the circular polarization does not match that of the helix (aGST-RCP), less pronounced currents will also be observed in the helices because of coupling. As expected, in the case of a low CD at 900 nm, there is minimal interaction under LCP [Fig. 3f] and RCP [Fig. 3g], where the extinction coefficient of amorphous and crystalline GST is lower.
Moreover, a rather interesting current distribution pattern is observed along the surface of the partially conductive rod where converging (red) and diverging (blue) points are formed, tracing out a rotating wave with a wavelength equal to that of the polarized plane wave source. 

\subsection{Angular resolved Mueller Matrix spectroscopic ellipsometry}

\begin{figure}[ht!]
\centering
\includegraphics[width=1.0\linewidth]{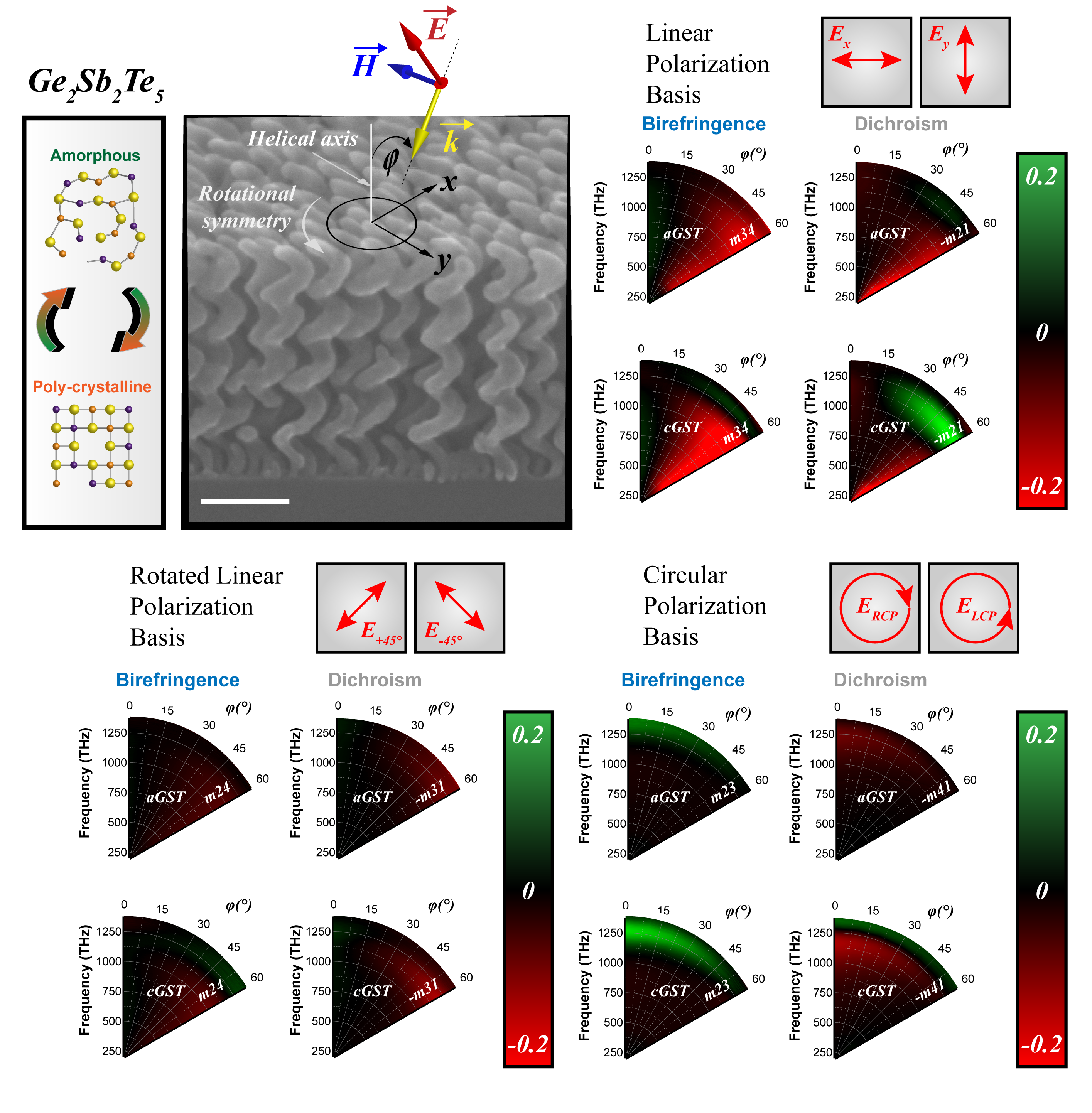}
\caption{Birefringence and dichorism of aGST and cGST in three unique polarization basis pairs measured using transmissive angular resolved Mueller Matrix spectroscopic ellipsometry}
\label{fig:current}
\end{figure}

In this subsection, spectral results when the AOI (angle of incidence) is tuned from normal incidence to 70$^\circ$ are reported. Since the GST chiral PCNMs possess in-plane rotational symmetry, when displaced on a transparent substrate, one can arbitrarily select $\theta$ and sweep the angle of elevation defined as $\phi$ as shown in SEM in Fig. \ref{fig:current}. The Mueller Matrix measurement technique is over-specified, which results in several redundancies observed in the Mueller Matrix results. As such, only 6 elements need to be extracted to analyze the birefringence and diattenuation (i.e. dichroism) for the intensities of the three polarization basis pairs displayed in Fig. \ref{fig:current}. 

There is no noticeable spectral depolarization in the medium when $\phi \geq 40^\circ$. As shown in the previous subsection for normal incidence measurements, a strong CB peak exists at the CD crossover point. Interesting enough, optical activity, manifested in the CB and CD polar plots, remains stable up until $\phi=40^\circ$, in which the medium begins to exhibit strong linear birefringence and dichorism at longer wavelengths. The signals are relatively weak in the amorphous and crystalline states for the rotated LP basis set. Moreover, the right-handed structure exhibits a negative cotton effect. 
\begin{figure*}[ht!]
\centering
\includegraphics[width=0.5\linewidth]{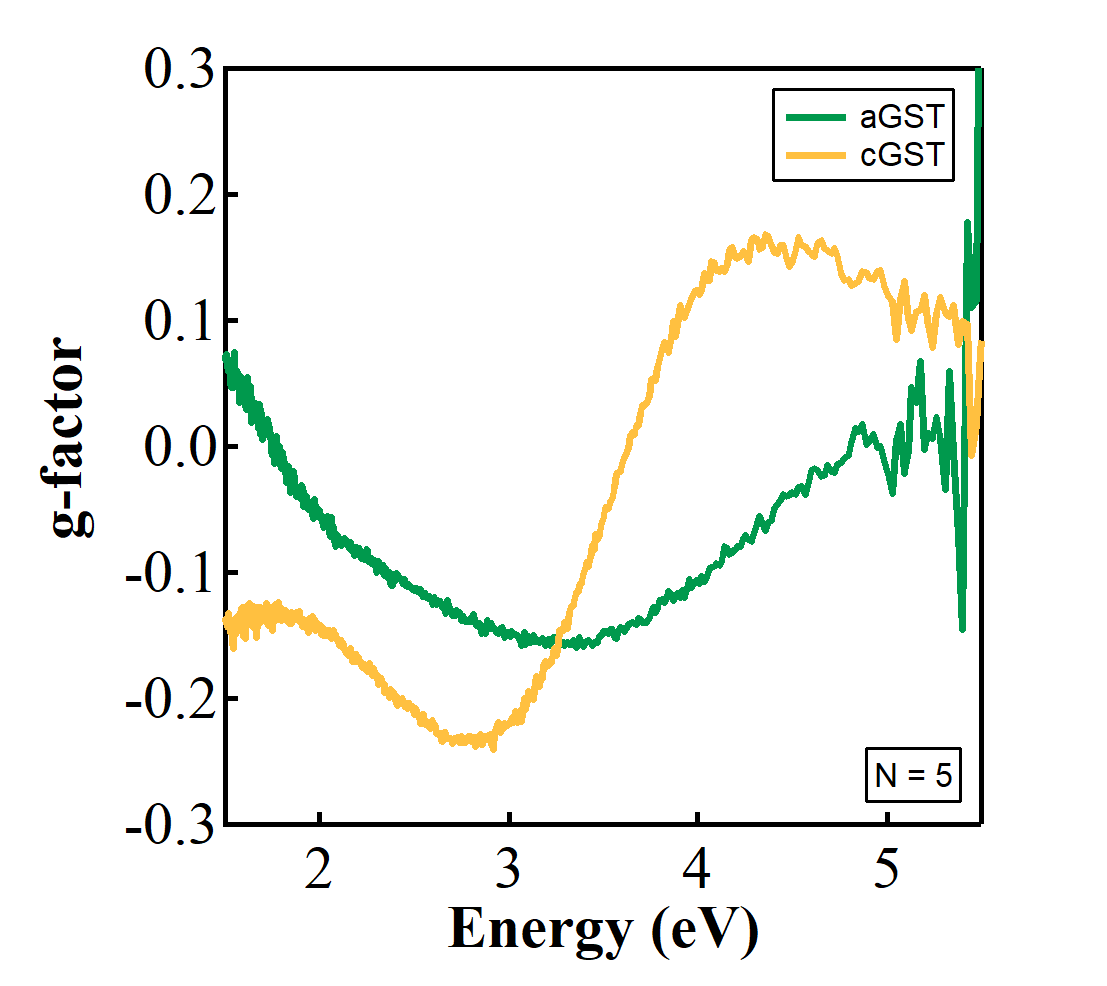}
\caption{g-factor for amorphous and crystalline states for a polar angle $\phi=20^{^\circ}$.} 
\label{fig:gfactor}
\end{figure*}
Angular resolved Mueller Matrix measurements in reflection mode were also performed to calculate the g-factor in terms of Mueller Matrix elements\cite{KilicAFM2021} as
\begin{align*} \text{g-factor} &= 2\dfrac{(A_{L}-A_{R})}{(A_{L}+A_{R})} = 2\dfrac{M^{T}_{14}+M^{R}_{14}}{1-M^{T}_{11}-M^{R}_{11}}.
\end{align*}
In the g-factor spectra measured at $\phi=20^{\circ}$, a broadband response is observed in the amorphous state where the strongest response is around 0.1; however, there is an increase in the g-factor upon crystallization and a reversal point coincides with the maximum circular birefringence response. 

\section{Reversible opto-thermal operation}

\subsection{Single-pulse transient transmission characterization}

\begin{figure}[ht!]
\centering
\includegraphics[width=1\linewidth]{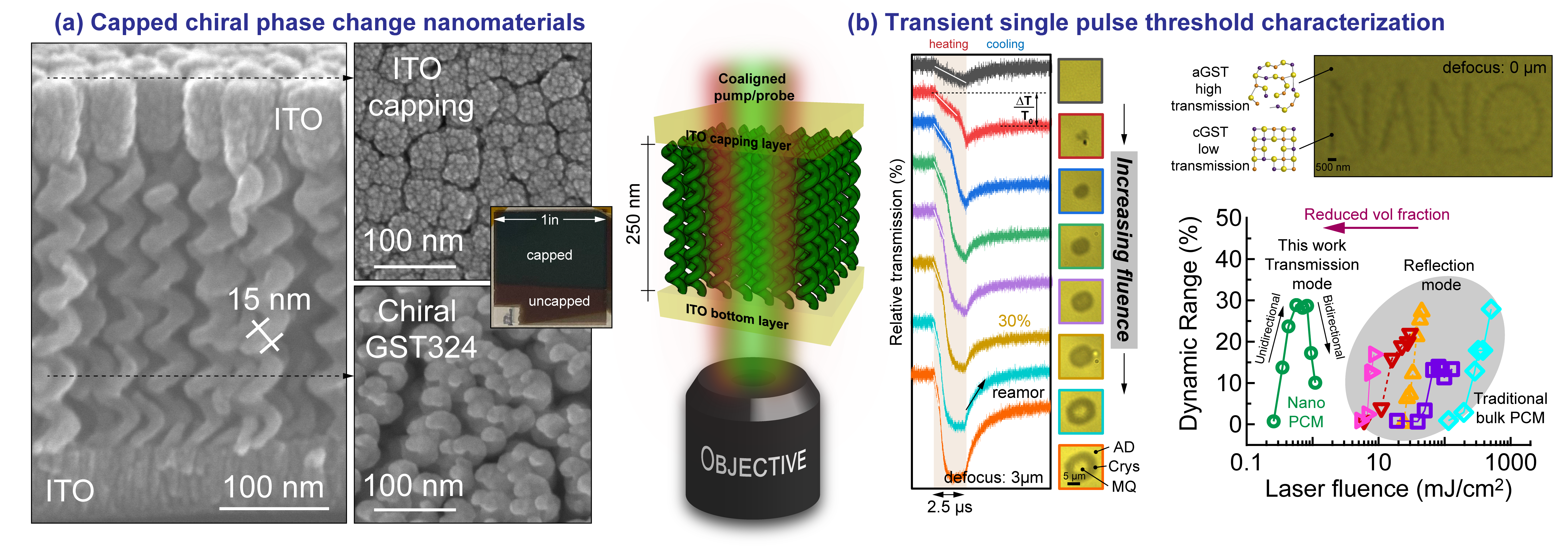}
\caption{(a) Cross-section and and top view SEMs of ITO sandwiched chiral PCNMs (b) Transient single-pulse threshold characterization comparing PCNMs to traditional bulk thin films}
\label{fig:singlepulse}
\end{figure}

The chiral PCNM medium must be capped with a thin layer of material to prevent GST from evaporating during the melt-quench process to cycle between amorphous and metastable crystalline states. The capping layer, consisting of sputtered indium tin oxide (ITO), also assists with heat dissipation properties as it is reported to have a thermal conductivity of 10 W/mK, twenty times that of GST. To further improve the thermal response, the chiral medium is deposited on a transparent glass substrate featuring a second sputter layer (provided commercially by 2spi) 
that reports a refractive index of about 2+0.004i in the visible regime. The capping layer is grown in-house from an ITO target at room temperature directly after the GLAD growth of GST chiral nanorods. During the capping layer deposition, the ITO condenses on the individual tips of the chiral PCNMs forming a semi-porous medium of ITO clusters as shown in Fig \ref{fig:singlepulse}(a).

The encapsulated samples are mounted in a custom-built optical microscopy setup featuring a 532 nm switching laser and a coaligned 632 nm read Gaussian beam of which are both focused onto a substrate using a 100X 1.3 NA oil immersion objective. A backward propagating incoherent white light halogen source is used to image the surface of the sample. The sample is mounted onto a closed-loop XYZ piezo stage allowing for precise spatial control for patterning applications. The intensity of the continuous-wave 532 nm fundamental lasing mode is varied in time using an amplitude modulator where the electrical signal is encoded by a multi-channel, high-speed pulse delay generator. This feature in the optical setup allows for precise control of the pulse duration and amplitude for programmable sequences. 

A series of optical transient threshold experiments are first conducted to identify the pulse dynamics required to achieve controlled reversible cycles between crystallization and amorphization states of the chalcogenide medium. Using a focused beam with a spot size of 400 nm, the word ``NANO" is optically written into the region where dark regions correspond to crystalline GST. For single pulse threshold testing, the coaligned beams are defocused by 3 $\mu$m steps, and the fluence of a 532 nm single-pulse is monotonically increased to determine the opto-thermal threshold parameters. In Fig 6(b), the transient transmission for a 2.5 $\mu$s pulse is plotted for increasing powers. The respective bright-field images are displayed, indicating a crystallization region with high extinction and good consistency with the transient measurements. The dynamic range saturates near 300 nJ/cm$^2$ before exhibiting amorphization in the most intense region of the Gaussian beam. Overall, the PCNMs exhibits lower fluences for switching due to the reduction in volume when compared to traditional bulk thin GST demonstrations\cite{SevisonACS,SinglePulsePLD_GST_AFM,Sun2016,Arjunan2020_ACS_E_materiasl}. While the volume fraction is reduced, the dynamic range of crystallization is comparable with bulk thin-film materials but requires much less energy, as exhibited in the opto-thermal threshold testing. 

\subsection{Chirality preservation}
To confirm the chiral structural preservation and tunable response, spot-by-spot crystallization and re-amorphization is employed to create a measurement area of 160 $\times$ 160 $\mu$m$^2$ which is sufficient for reflective Mueller Matrix measurements at a $\phi=55^{\circ}$ incident angle. First, one region is crystallized, then a second region is crystallized, followed by reamorphous single pulses using appropriate opto-thermal pulse characteristics. The regions are then measured on the Mueller Matrix spectroscopic ellipsometer, and the resulting m$_{23}$ and m$_{14}$ elements are plotted in Fig \ref{fig:chiralpreservation}(a). As-deposited (AD) chiral PCNMs exhibit a noticeable peak in the m$_{23}$ spectra which is proportional to circular birefringence at 2.5 eV, which increases in magnitude upon crystallization. Moreover, the medium exhibits a circular dichroism response at 2 eV. Upon crystallization, the resonance is doubled and reversed back to the AD state after the material has been melt-quenched (The slight residual difference in CD can be attributed to the standard difference between the pristine AD and the MQ states). Notably, the non-zero m$_{23}$ and m$_{14}$ features indicate that the structural properties are indeed preserved (Fig \ref{fig:chiralpreservation}(b)). 
\begin{figure}[ht!]
\centering
\includegraphics[width=1.0\linewidth]{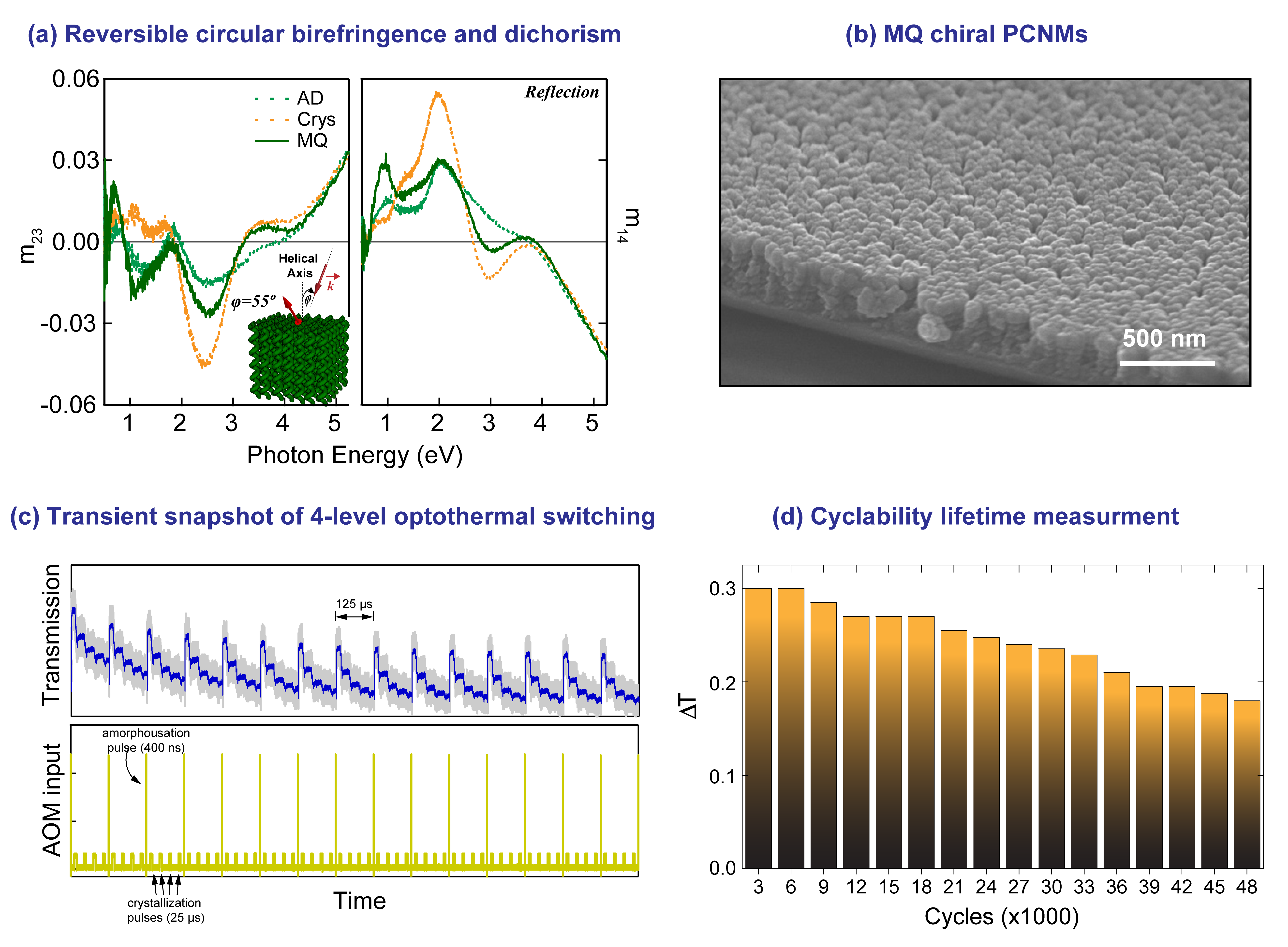}
\caption{(a) Mueller Matrix elements of as-deposited (AD), crystallized (Crys), and melt-quench (MQ) intrinsically chiral GST PCNMs, (b) SEM of cycled GST PCNMs, (c) Snapshot of lifetime measurements, and (d) Transmission difference as a function of number of cycles.}
\label{fig:chiralpreservation}
\end{figure}

\subsection{Optical switching material lifetime measurements}
Opto-thermal material lifetime measurements are also carried out, where a crystallization and amorphization programming sequence consisting of a series of 25 $\mu$s pulses followed by one 400 ns pulse as shown in Fig. 7c are employed. 
The 532 nm Gaussian beam is defocused
by 3 $\mu$m to improve the uniformity of the heat distribution in the switching region, which
provides the switching reliability and prevents undesirable “hot-spots”. As shown in Fig. 7c, the transmission contrast upon cycling is perfectly preserved over the first 15 cycles, with appreciable contrast still maintained even after 50,000 cycles (Fig 7d). Additional cycles are shown in Supplemental Note S6. Crucially, this structural integrity of the chiral PCNMs after thousands of cycles of melt-quench/crystallization shows their promise as a robust solid-state platform for dynamic polarization control at the nanoscale. In fact, as a preliminary demonstration, an electrically addressed micro-pixel based on the PCNMs platform is shown in Supplemental Note 7 and the supplemental video with an integrated planar microheater providing uniform switching over $7 \times 7~\mu$m$^2$ area.


\section{Conclusion}

In summary, intrinsically chiral nanorods consisting of phase change material GST exhibiting optical activity are fabricated, experimentally characterized, and numerically investigated. The angular resolved Mueller Matrix method measured differential absorption of left and right circularly polarized light components and circular birefringence. Additionally, endurance testing shows the stability of the platform up to 50,000 cycles of amorphization/crystallization. As discussed, the amorphous and crystalline stable phase states of GST offer a novel route for tunability due to the stark difference in the complex refractive index where selective material absorption between circular eigenpolarization states is the driving factor in this scheme. The research outlined here has led to a much better understanding of the material properties that affect the optical performance of chalcogenides and illustrates an alternative bottom-up growth approach for developing high-performance dual-mode electro-optic devices targeted for the visible regime and beyond. The results of this work could be pivotal in realizing large area, reconfigurable chalcogenide optical devices with high endurance and longer lifetimes. 
 

\section{Methods}
\subsection{Nanomaterial fabrication}
Chiral PCNMs are synthesized via the GLAD technique using a MDC evap4000 e-beam evaporator. The substrates are mounted 35 cm above the evaporant source at a fixed deposition angle ($\alpha$) while a voltage-controlled motor enables continuous rotation at pre-selected angular speeds ranging between 2-9 revolutions per ms about the surface normal in a CW or ACW fashion. The chamber base pressure is kept around $<$1 $\mu$T with minimal increase in pressure during deposition. In addition, a QCM is enabled for precise control over the deposition rate. For all films, the deposition material Ge$_2$Sb$_2$Te$_5$ (99.99\% pellets Plasmaterials) is evaporated at relatively low e-beam filament currents (2-4 mA) when compared to transition metals and dielectrics, presumably due to the low vapor pressure of Te in a Te dominated glass composition. 

\subsection{Finite-difference time-domain (FDTD) simulations}
Finite-difference-time-domain (FDTD) full-wave analysis numerical solver (Lumerical, ANSYS) is implemented to calculate the resulting transmittance, reflectance, and absorptance from left and right-handed circularly polarized light. 
The nanostructure array is simulated using periodic boundary conditions. Perfectly matched layer BCs is assigned to the input and output ports. Two orthogonal plane wave sources are assigned that are out of phase by $\pm$ 90$^{\circ}$ to implement circularly polarized light excitations. Absolute and relative (transmissive dissymmetry) CD are calculated from independent calculations for LCP and RCP excitations. 
\subsection{In-situ optical measurements}
Reversible optical crystallization is achieved using a 532 nm diode laser that is modulated by a programmable pulse generator to produce optical pulses of varying time durations with a fixed 5 ns decay time. A coaligned continuous wave 633 nm HeNe laser is focused using a 1.3 NA objective with 100X magnification to create sub-micron spot sizes with high fluences for reversible switching experiments. The maximum input power of the 532 nm laser to the focusing objective is roughly 65 mW for an 8V input into the AOM. A 50X Mitutoyo Plan Apo infinity corrected, 0.42 numerical aperture objective is used to collect the transmitted beams and for transmissive imaging on a CCD camera from a backward propagating white light source. The sample is mounted onto a 3D piezo stage (Newport) to control the XYZ position precisely. For reliable crystallization, a slightly defocused spot is used, which ensures no damage in the center of the spot.

\subsection{In-situ electrical switching measurements}
The microstrip sample is created by spot-by-spot ablation using the 532 nm laser with high fluence to remove the ITO/chiral PCNMs/ITO layers to altogether. This maskless approach patterning approach allows for one to create an electrically isolated pixel. High-speed probe tips are placed onto the bottom ITO layer by digging through the fragile capped nanostructured layer. A single 8V 25 $\mu$s pulse is sent through the ITO for localized heating in a 7 $\times$ 7 $\mu$m$^2$. 
\begin{acknowledgement}

The authors thank Professor Qiwen Zhan for fruitful discussion and guidance. This work was supported in part by the National Science Foundation (Collaborative Research, Grant No 1710273) and Thorlabs Inc. J.A.B. gratefully acknowledges financial support from the National Academies under the Ford Foundation Fellowship Program.  T.A.S. is supported by the ONR HBCU/MI Program: N00014-20-1-2541 and acknowledges support from the Martin Luther King Visiting Scholars Program at MIT. 
\end{acknowledgement}


\section{Data availability}

The data that support the findings of this study are available from the corresponding authors upon reasonable request.


\bibliography{achemso}

\providecommand{\latin}[1]{#1}
\makeatletter
\providecommand{\doi}
  {\begingroup\let\do\@makeother\dospecials
  \catcode`\{=1 \catcode`\}=2 \doi@aux}
\providecommand{\doi@aux}[1]{\endgroup\texttt{#1}}
\makeatother
\providecommand*\mcitethebibliography{\thebibliography}
\csname @ifundefined\endcsname{endmcitethebibliography}
  {\let\endmcitethebibliography\endthebibliography}{}
\begin{mcitethebibliography}{34}
\providecommand*\natexlab[1]{#1}
\providecommand*\mciteSetBstSublistMode[1]{}
\providecommand*\mciteSetBstMaxWidthForm[2]{}
\providecommand*\mciteBstWouldAddEndPuncttrue
  {\def\EndOfBibitem{\unskip.}}
\providecommand*\mciteBstWouldAddEndPunctfalse
  {\let\EndOfBibitem\relax}
\providecommand*\mciteSetBstMidEndSepPunct[3]{}
\providecommand*\mciteSetBstSublistLabelBeginEnd[3]{}
\providecommand*\EndOfBibitem{}
\mciteSetBstSublistMode{f}
\mciteSetBstMaxWidthForm{subitem}{(\alph{mcitesubitemcount})}
\mciteSetBstSublistLabelBeginEnd
  {\mcitemaxwidthsubitemform\space}
  {\relax}
  {\relax}

\bibitem[Qiu \latin{et~al.}(2018)Qiu, \latin{et~al.}
  others]{Qui2018IntrinsicChiralityReview}
Qiu,~M., \latin{et~al.}  3D Metaphotonic Nanostructures with Intrinsic
  Chirality. \emph{Advanced Functional Materials} \textbf{2018}, \emph{28},
  1803147\relax
\mciteBstWouldAddEndPuncttrue
\mciteSetBstMidEndSepPunct{\mcitedefaultmidpunct}
{\mcitedefaultendpunct}{\mcitedefaultseppunct}\relax
\EndOfBibitem
\bibitem[Mun \latin{et~al.}(2020)Mun, \latin{et~al.} others]{Mun2020}
Mun,~J., \latin{et~al.}  Electromagnetic chirality: from fundamentals to
  nontraditional chiroptical phenomena. \emph{Light: Science {\&} Applications}
  \textbf{2020}, \emph{9}, 139\relax
\mciteBstWouldAddEndPuncttrue
\mciteSetBstMidEndSepPunct{\mcitedefaultmidpunct}
{\mcitedefaultendpunct}{\mcitedefaultseppunct}\relax
\EndOfBibitem
\bibitem[Liu \latin{et~al.}(2020)Liu, \latin{et~al.} others]{StereoMeta20_AM}
Liu,~Z., \latin{et~al.}  Fano-Enhanced Circular Dichroism in Deformable Stereo
  Metasurfaces. \emph{Advanced Materials} \textbf{2020}, \emph{32},
  1907077\relax
\mciteBstWouldAddEndPuncttrue
\mciteSetBstMidEndSepPunct{\mcitedefaultmidpunct}
{\mcitedefaultendpunct}{\mcitedefaultseppunct}\relax
\EndOfBibitem
\bibitem[Zhao \latin{et~al.}(2012)Zhao, Belkin, and Al{\`u}]{Zhao2012}
Zhao,~Y.; Belkin,~M.~A.; Al{\`u},~A. Twisted optical metamaterials for
  planarized ultrathin broadband circular polarizers. \emph{Nature
  Communications} \textbf{2012}, \emph{3}, 870\relax
\mciteBstWouldAddEndPuncttrue
\mciteSetBstMidEndSepPunct{\mcitedefaultmidpunct}
{\mcitedefaultendpunct}{\mcitedefaultseppunct}\relax
\EndOfBibitem
\bibitem[Esposito \latin{et~al.}(2015)Esposito, \latin{et~al.}
  others]{plamonicsCD}
Esposito,~M., \latin{et~al.}  Nanoscale 3D Chiral Plasmonic Helices with
  Circular Dichroism at Visible Frequencies. \emph{ACS Photonics}
  \textbf{2015}, \emph{2}, 105--114\relax
\mciteBstWouldAddEndPuncttrue
\mciteSetBstMidEndSepPunct{\mcitedefaultmidpunct}
{\mcitedefaultendpunct}{\mcitedefaultseppunct}\relax
\EndOfBibitem
\bibitem[Dietrich \latin{et~al.}(2012)Dietrich, Lehr, Helgert, Tünnermann, and
  Kley]{plasmonicAOM}
Dietrich,~K.; Lehr,~D.; Helgert,~C.; Tünnermann,~A.; Kley,~E.-B. Circular
  Dichroism from Chiral Nanomaterial Fabricated by On-Edge Lithography.
  \emph{Advanced Materials} \textbf{2012}, \emph{24}, OP321--OP325\relax
\mciteBstWouldAddEndPuncttrue
\mciteSetBstMidEndSepPunct{\mcitedefaultmidpunct}
{\mcitedefaultendpunct}{\mcitedefaultseppunct}\relax
\EndOfBibitem
\bibitem[Krause and Brett(2008)Krause, and Brett]{GradientAFM08}
Krause,~K.~M.; Brett,~M.~J. Spatially Graded Nanostructured Chiral Films as
  Tunable Circular Polarizers. \emph{Advanced Functional Materials}
  \textbf{2008}, \emph{18}, 3111--3118\relax
\mciteBstWouldAddEndPuncttrue
\mciteSetBstMidEndSepPunct{\mcitedefaultmidpunct}
{\mcitedefaultendpunct}{\mcitedefaultseppunct}\relax
\EndOfBibitem
\bibitem[Zhu \latin{et~al.}(2018)Zhu, \latin{et~al.} others]{Zhu2018}
Zhu,~A.~Y., \latin{et~al.}  Giant intrinsic chiro-optical activity in planar
  dielectric nanostructures. \emph{Light: Science {\&} Applications}
  \textbf{2018}, \emph{7}, 17158--17158\relax
\mciteBstWouldAddEndPuncttrue
\mciteSetBstMidEndSepPunct{\mcitedefaultmidpunct}
{\mcitedefaultendpunct}{\mcitedefaultseppunct}\relax
\EndOfBibitem
\bibitem[Li \latin{et~al.}(2021)Li, \latin{et~al.}
  others]{TunableDielectricsAlu21}
Li,~J., \latin{et~al.}  Tunable Chiral Optics in All-Solid-Phase Reconfigurable
  Dielectric Nanostructures. \emph{Nano Letters} \textbf{2021}, \emph{21},
  973--979, PMID: 33372805\relax
\mciteBstWouldAddEndPuncttrue
\mciteSetBstMidEndSepPunct{\mcitedefaultmidpunct}
{\mcitedefaultendpunct}{\mcitedefaultseppunct}\relax
\EndOfBibitem
\bibitem[de~Galarreta \latin{et~al.}(2020)de~Galarreta, \latin{et~al.}
  others]{RuizdeGalarreta:20}
de~Galarreta,~C.~R., \latin{et~al.}  Reconfigurable multilevel control of
  hybrid all-dielectric phase-change metasurfaces. \emph{Optica} \textbf{2020},
  \emph{7}, 476--484\relax
\mciteBstWouldAddEndPuncttrue
\mciteSetBstMidEndSepPunct{\mcitedefaultmidpunct}
{\mcitedefaultendpunct}{\mcitedefaultseppunct}\relax
\EndOfBibitem
\bibitem[Abdollahramezani \latin{et~al.}(2021)Abdollahramezani, \latin{et~al.}
  others]{PopNanoLetts21}
Abdollahramezani,~S., \latin{et~al.}  Dynamic Hybrid Metasurfaces. \emph{Nano
  Letters} \textbf{2021}, \emph{21}, 1238--1245, PMID: 33481600\relax
\mciteBstWouldAddEndPuncttrue
\mciteSetBstMidEndSepPunct{\mcitedefaultmidpunct}
{\mcitedefaultendpunct}{\mcitedefaultseppunct}\relax
\EndOfBibitem
\bibitem[de~Galarreta \latin{et~al.}(2018)de~Galarreta, \latin{et~al.}
  others]{GalarretaBeamSteering}
de~Galarreta,~C.~R., \latin{et~al.}  Nonvolatile Reconfigurable Phase-Change
  Metadevices for Beam Steering in the Near Infrared. \emph{Advanced Functional
  Materials} \textbf{2018}, \emph{28}, 1704993\relax
\mciteBstWouldAddEndPuncttrue
\mciteSetBstMidEndSepPunct{\mcitedefaultmidpunct}
{\mcitedefaultendpunct}{\mcitedefaultseppunct}\relax
\EndOfBibitem
\bibitem[Zhang \latin{et~al.}(2021)Zhang, Fowler, Liang, Azhar, Shalaginov,
  Deckoff-Jones, An, Chou, Roberts, Liberman, Kang, R{\'i}os, Richardson,
  Rivero-Baleine, Gu, Zhang, and Hu]{Zhang2021}
Zhang,~Y. \latin{et~al.}  Electrically reconfigurable non-volatile metasurface
  using low-loss optical phase-change material. \emph{Nature Nanotechnology}
  \textbf{2021}, \relax
\mciteBstWouldAddEndPunctfalse
\mciteSetBstMidEndSepPunct{\mcitedefaultmidpunct}
{}{\mcitedefaultseppunct}\relax
\EndOfBibitem
\bibitem[R{\'i}os \latin{et~al.}(2016)R{\'i}os, Hosseini, Taylor, and
  Bhaskaran]{RiosAM}
R{\'i}os,~C.; Hosseini,~P.; Taylor,~R.~A.; Bhaskaran,~H. Color Depth Modulation
  and Resolution in Phase-Change Material Nanodisplays. \emph{Advanced
  Materials} \textbf{2016}, \emph{28}, 4720--4726\relax
\mciteBstWouldAddEndPuncttrue
\mciteSetBstMidEndSepPunct{\mcitedefaultmidpunct}
{\mcitedefaultendpunct}{\mcitedefaultseppunct}\relax
\EndOfBibitem
\bibitem[R{\'i}os \latin{et~al.}(2015)R{\'i}os, \latin{et~al.}
  others]{Rios2015}
R{\'i}os,~C., \latin{et~al.}  Integrated all-photonic non-volatile multi-level
  memory. \emph{Nature Photonics} \textbf{2015}, \emph{9}, 725--732\relax
\mciteBstWouldAddEndPuncttrue
\mciteSetBstMidEndSepPunct{\mcitedefaultmidpunct}
{\mcitedefaultendpunct}{\mcitedefaultseppunct}\relax
\EndOfBibitem
\bibitem[Cao \latin{et~al.}(2013)Cao, Zhang, Simpson, Wei, and Cryan]{Cao13}
Cao,~T.; Zhang,~L.; Simpson,~R.~E.; Wei,~C.; Cryan,~M.~J. Strongly tunable
  circular dichroism in gammadion chiral phase-change metamaterials. \emph{Opt.
  Express} \textbf{2013}, \emph{21}, 27841--27851\relax
\mciteBstWouldAddEndPuncttrue
\mciteSetBstMidEndSepPunct{\mcitedefaultmidpunct}
{\mcitedefaultendpunct}{\mcitedefaultseppunct}\relax
\EndOfBibitem
\bibitem[Ding \latin{et~al.}(2021)Ding, Rui, Gu, Zhan, and Cui]{Ding21}
Ding,~C.; Rui,~G.; Gu,~B.; Zhan,~Q.; Cui,~Y. Phase-change metasurface with
  tunable and switchable circular dichroism. \emph{Opt. Lett.} \textbf{2021},
  \emph{46}, 2525--2528\relax
\mciteBstWouldAddEndPuncttrue
\mciteSetBstMidEndSepPunct{\mcitedefaultmidpunct}
{\mcitedefaultendpunct}{\mcitedefaultseppunct}\relax
\EndOfBibitem
\bibitem[Yin \latin{et~al.}(2015)Yin, Schäferling, Michel, Tittl, Wuttig,
  Taubner, and Giessen]{Yin2015ActiveChiralPlasmonics}
Yin,~X.; Schäferling,~M.; Michel,~A.-K.~U.; Tittl,~A.; Wuttig,~M.;
  Taubner,~T.; Giessen,~H. Active Chiral Plasmonics. \emph{Nano Letters}
  \textbf{2015}, \emph{15}, 4255--4260, PMID: 26039735\relax
\mciteBstWouldAddEndPuncttrue
\mciteSetBstMidEndSepPunct{\mcitedefaultmidpunct}
{\mcitedefaultendpunct}{\mcitedefaultseppunct}\relax
\EndOfBibitem
\bibitem[Shanmugam \latin{et~al.}(2019)Shanmugam, \latin{et~al.}
  others]{PhotoInducedGST2019}
Shanmugam,~J., \latin{et~al.}  Giant Photoinduced Chirality in Thin Film
  Ge$_2$Sb$_2$Te$_5$. \emph{physica status solidi (RRL) – Rapid Research
  Letters} \textbf{2019}, \emph{13}, 1900449\relax
\mciteBstWouldAddEndPuncttrue
\mciteSetBstMidEndSepPunct{\mcitedefaultmidpunct}
{\mcitedefaultendpunct}{\mcitedefaultseppunct}\relax
\EndOfBibitem
\bibitem[Borisenko \latin{et~al.}(2015)Borisenko, \latin{et~al.}
  others]{Borisenko2015}
Borisenko,~K.~B., \latin{et~al.}  Photo-induced optical activity in
  phase-change memory materials. \emph{Scientific Reports} \textbf{2015},
  \emph{5}, 8770\relax
\mciteBstWouldAddEndPuncttrue
\mciteSetBstMidEndSepPunct{\mcitedefaultmidpunct}
{\mcitedefaultendpunct}{\mcitedefaultseppunct}\relax
\EndOfBibitem
\bibitem[Hawkeye \latin{et~al.}(2014)Hawkeye, Taschuk, and Brett]{hawkeye14}
Hawkeye,~M.~M.; Taschuk,~M.~T.; Brett,~M.~J. \emph{Glancing Angle Deposition of
  Thin Films: Engineering the Nanoscale}; Wiley: London, 2014\relax
\mciteBstWouldAddEndPuncttrue
\mciteSetBstMidEndSepPunct{\mcitedefaultmidpunct}
{\mcitedefaultendpunct}{\mcitedefaultseppunct}\relax
\EndOfBibitem
\bibitem[Starbova \latin{et~al.}(1997)Starbova, Dikova, and
  Starbov]{STARBOVA1997261_As2S3}
Starbova,~K.; Dikova,~J.; Starbov,~N. Structure related properties of obliquely
  deposited amorphous a-As2S3 thin films. \emph{Journal of Non-Crystalline
  Solids} \textbf{1997}, \emph{210}, 261--266\relax
\mciteBstWouldAddEndPuncttrue
\mciteSetBstMidEndSepPunct{\mcitedefaultmidpunct}
{\mcitedefaultendpunct}{\mcitedefaultseppunct}\relax
\EndOfBibitem
\bibitem[Bhardwaj \latin{et~al.}(2007)Bhardwaj, Shishodia, and
  Mehra]{Bhardwaj2007GeSe2}
Bhardwaj,~P.; Shishodia,~P.~K.; Mehra,~R.~M. Optical and electrical properties
  of obliquely deposited a-GeSe2 films. \emph{Journal of Materials Science}
  \textbf{2007}, \emph{42}, 1196--1201\relax
\mciteBstWouldAddEndPuncttrue
\mciteSetBstMidEndSepPunct{\mcitedefaultmidpunct}
{\mcitedefaultendpunct}{\mcitedefaultseppunct}\relax
\EndOfBibitem
\bibitem[Martín-Palma \latin{et~al.}(2007)Martín-Palma, Ryan, and
  Pantano]{GSSnanorods}
Martín-Palma,~R.~J.; Ryan,~J.~V.; Pantano,~C.~G. Spectral behavior of the
  optical constants in the visible\/near infrared of GeSbSe chalcogenide thin
  films grown at glancing angle. \emph{Journal of Vacuum Science \& Technology
  A} \textbf{2007}, \emph{25}, 587--591\relax
\mciteBstWouldAddEndPuncttrue
\mciteSetBstMidEndSepPunct{\mcitedefaultmidpunct}
{\mcitedefaultendpunct}{\mcitedefaultseppunct}\relax
\EndOfBibitem
\bibitem[Zhang \latin{et~al.}(2021)Zhang, \latin{et~al.}
  others]{ZhangeAPLPerspectives2021}
Zhang,~Y., \latin{et~al.}  Myths and truths about optical phase change
  materials: A perspective. \emph{Applied Physics Letters} \textbf{2021},
  \emph{118}, 210501\relax
\mciteBstWouldAddEndPuncttrue
\mciteSetBstMidEndSepPunct{\mcitedefaultmidpunct}
{\mcitedefaultendpunct}{\mcitedefaultseppunct}\relax
\EndOfBibitem
\bibitem[Guo \latin{et~al.}(2018)Guo, \latin{et~al.} others]{GuoNidoped}
Guo,~P., \latin{et~al.}  Improving the performance of Ge$_2$Sb$_2$Te$_5$
  materials via nickel doping: Towards RF-compatible phase-change devices.
  \emph{Applied Physics Letters} \textbf{2018}, \emph{113}, 171903\relax
\mciteBstWouldAddEndPuncttrue
\mciteSetBstMidEndSepPunct{\mcitedefaultmidpunct}
{\mcitedefaultendpunct}{\mcitedefaultseppunct}\relax
\EndOfBibitem
\bibitem[Guo \latin{et~al.}(2020)Guo, \latin{et~al.} others]{GuoWdoped}
Guo,~P., \latin{et~al.}  Tungsten-doped Ge$_2$Sb$_2$Te$_5$ phase change
  material for high-speed optical switching devices. \emph{Applied Physics
  Letters} \textbf{2020}, \emph{116}, 131901\relax
\mciteBstWouldAddEndPuncttrue
\mciteSetBstMidEndSepPunct{\mcitedefaultmidpunct}
{\mcitedefaultendpunct}{\mcitedefaultseppunct}\relax
\EndOfBibitem
\bibitem[Mukherjee and Gall(2013)Mukherjee, and Gall]{MUKHERJEE2013158}
Mukherjee,~S.; Gall,~D. Structure zone model for extreme shadowing conditions.
  \emph{Thin Solid Films} \textbf{2013}, \emph{527}, 158--163\relax
\mciteBstWouldAddEndPuncttrue
\mciteSetBstMidEndSepPunct{\mcitedefaultmidpunct}
{\mcitedefaultendpunct}{\mcitedefaultseppunct}\relax
\EndOfBibitem
\bibitem[Kilic \latin{et~al.}(2021)Kilic, \latin{et~al.} others]{KilicAFM2021}
Kilic,~U., \latin{et~al.}  Broadband Enhanced Chirality with Tunable Response
  in Hybrid Plasmonic Helical Metamaterials. \emph{Advanced Functional
  Materials} \textbf{2021}, \emph{31}, 2010329\relax
\mciteBstWouldAddEndPuncttrue
\mciteSetBstMidEndSepPunct{\mcitedefaultmidpunct}
{\mcitedefaultendpunct}{\mcitedefaultseppunct}\relax
\EndOfBibitem
\bibitem[Sevison \latin{et~al.}(2020)Sevison, \latin{et~al.}
  others]{SevisonACS}
Sevison,~G.~A., \latin{et~al.}  Phase Change Dynamics and Two-Dimensional 4-Bit
  Memory in Ge$_2$Sb$_2$Te$_5$ via Telecom-Band Encoding. \emph{ACS Photonics}
  \textbf{2020}, \emph{7}, 480--487\relax
\mciteBstWouldAddEndPuncttrue
\mciteSetBstMidEndSepPunct{\mcitedefaultmidpunct}
{\mcitedefaultendpunct}{\mcitedefaultseppunct}\relax
\EndOfBibitem
\bibitem[Lu \latin{et~al.}(2013)Lu, \latin{et~al.}
  others]{SinglePulsePLD_GST_AFM}
Lu,~H., \latin{et~al.}  Single Pulse Laser-Induced Phase Transitions of
  PLD-Deposited Ge$_2$Sb$_2$Te$_5$ Films. \emph{Advanced Functional Materials}
  \textbf{2013}, \emph{23}, 3621--3627\relax
\mciteBstWouldAddEndPuncttrue
\mciteSetBstMidEndSepPunct{\mcitedefaultmidpunct}
{\mcitedefaultendpunct}{\mcitedefaultseppunct}\relax
\EndOfBibitem
\bibitem[Sun \latin{et~al.}(2016)Sun, \latin{et~al.} others]{Sun2016}
Sun,~X., \latin{et~al.}  Crystallization of Ge$_2$Sb$_2$Te$_5$ thin films by
  nano- and femtosecond single laser pulse irradiation. \emph{Scientific
  Reports} \textbf{2016}, \emph{6}, 28246\relax
\mciteBstWouldAddEndPuncttrue
\mciteSetBstMidEndSepPunct{\mcitedefaultmidpunct}
{\mcitedefaultendpunct}{\mcitedefaultseppunct}\relax
\EndOfBibitem
\bibitem[Arjunan \latin{et~al.}(2020)Arjunan, \latin{et~al.}
  others]{Arjunan2020_ACS_E_materiasl}
Arjunan,~M.~S., \latin{et~al.}  Realization of 4-Bit Multilevel Optical
  Switching in Ge$_2$Sb$_2$Te$_5$ and Ag$_5$In$_5$Sb$_{60}$Te$_{30}$
  Phase-Change Materials Enabled in the Visible Region. \emph{ACS Applied
  Electronic Materials} \textbf{2020}, \emph{2}, 3977--3986\relax
\mciteBstWouldAddEndPuncttrue
\mciteSetBstMidEndSepPunct{\mcitedefaultmidpunct}
{\mcitedefaultendpunct}{\mcitedefaultseppunct}\relax
\EndOfBibitem
\end{mcitethebibliography}
\newpage

\begin{center}
 \section {Supplemental Information: \\Chiral Phase Change Nanomaterials}

\end{center}


\noindent \textbf{\large S1: E-beam evaporated GST at normal incidence}

\noindent \underline{Standard spectroscopic ellipsometry}\\
\noindent Prior to investigating the engineered PCNMs deposited by e-beam evaporation, we first characterize the optical properties of GST grown at normal incidence where a bulk thin film is produced. We employ ex situ standard spectroscopic ellipsometry on the resulting film to measure the change in polarization as varying wavelengths of light reflect from the surface of the GST films where the angle of incidence measured from the surface normal was set to 55$^\circ$. The reflected polarization change is represented by complex valued ratio, $\rho$ defined as
\begin{equation}
    \rho = \frac{r_{s}}{r_{p}} = \tan(\Psi)e^{i\Delta}
\end{equation}
\noindent where $r_{s}$ and $r_{p}$ are the Fresnel reflection coefficients for $s$ and $p$ polarized light, respectively, and the ellipsometric parameters $\Psi=\arctan(r_{p}/r_{s})$ and $\Delta=\delta_{p} - \delta_{s}$ are the relative amplitude and phase differences between the orthogonal polarization components. We then determine the optical parameters $n$ and $k$ using a Lorentz model with a regression algorithm where we employ the following merit function $M$ defined as
\begin{equation}
M = \mathlarger{\mathlarger{\sum}}_{i}^{N} X_i\Bigg[\bigg( \frac{\Delta_{i}-\Delta(\mathbf{x})}{\sigma_{\Delta}} \bigg)^{2} + \bigg( \frac{\Psi_{i}-\Psi(\mathbf{x})}{\sigma_{\Psi}} \bigg)^{2}\Bigg]^{2}\text{.}
\label{eq:refname1}
\end{equation}
The resulting refractive index $n$ and extinction coefficient $\kappa$ for amorphous and crystalline films are shown in Fig \ref{fig:nandk} evaporated films are plotted in dashed lines. During the fitting procedure we treated the material as a graded layer to obtain better fitting results, meaning that the real and imaginary permittivity increase from the bottom of the film to the top. We attribute the inhomogeneous nature of the resulting film is due to Ge-rich composition closer to the substrate because those molecues are lighter when compared with Sb and Te. Therefore, fractional molecules are more likely to evaporate first from the crucible.

\begin{figure*}[ht!]
\centering
\setcounter{figure}{0}    
\includegraphics[width=0.5\linewidth]{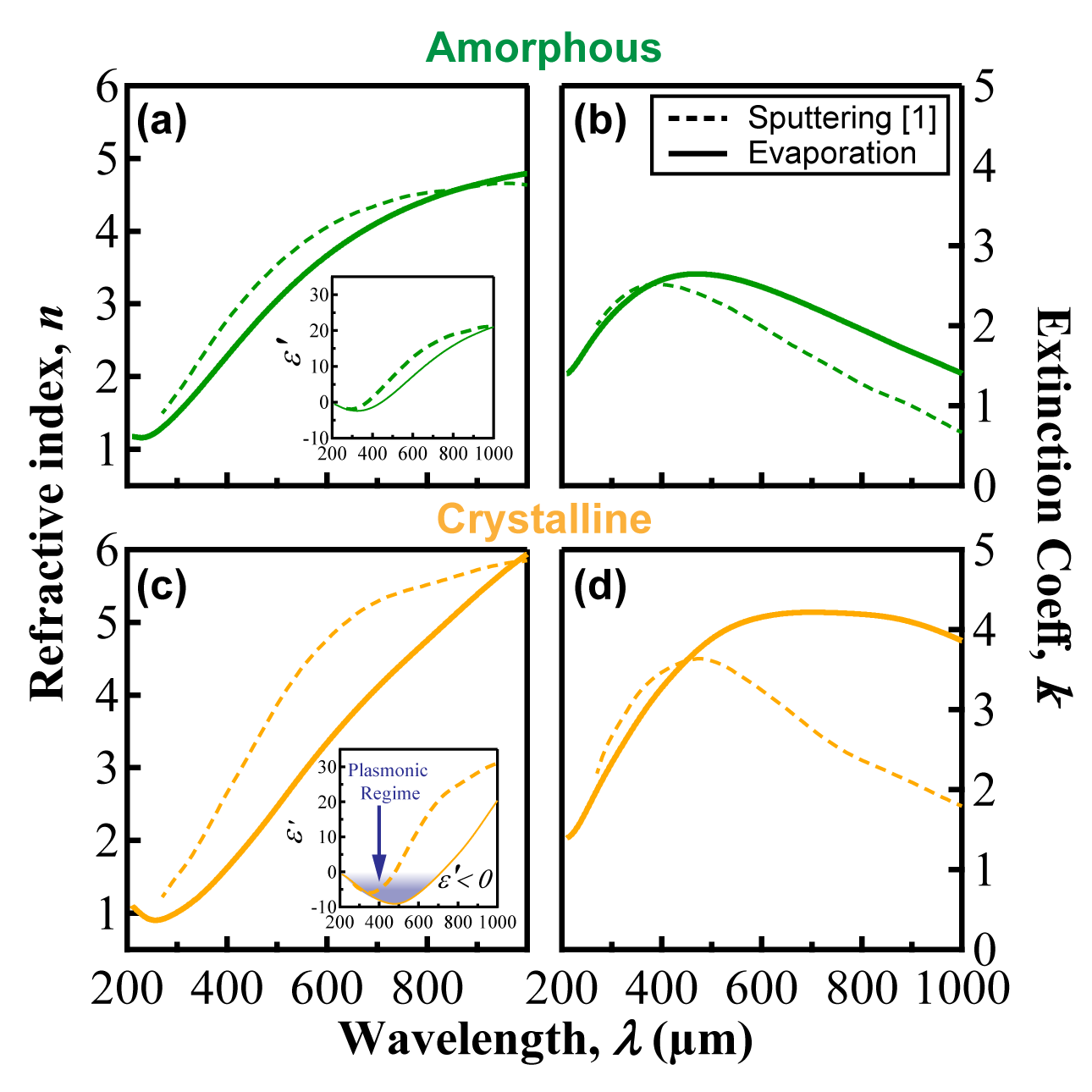}
\caption{Optical material properties of isotropic GST comparing RF-sputtering to e-beam evaporation growth techniques for aGST and cGST.} 
\label{fig:nandk}
\end{figure*}

This was found for the evaporative films for both the amorphous and crystalline case. Since optical properties of GST are known to vary depending on the deposition process, for context, we also plot the optical properties of sputtered films produced from a Ge$_2$Sb$_2$Te$_5$ target which are overlaid in solid lines. Additionally, we have plotted the dielectric constant $\epsilon_{1}$ of the material as well, indicating the plasmonic region of the semiconducting chalcogenide. The results presented are consistent with previously reported optical properties for germanium-antimony-tellurium based chalcogenides. GST prepared by evaporation shows smaller refractive index and larger extinction coefficient above 410 nm. This may be attributed to the fact evaporation is known to produce porous films when compared to the densely packed films deposited by conventional sputtering. Moreover, in a porous isotropic film, light will propagate at faster velocities and experience larger scattering losses in the medium.\\

\noindent \underline{Energy Dispersive Spectroscopy}\\
\noindent We perform Energy Dispersive X-Ray (EDX) Spectroscopy on an isotropic thin film deposited on a Si substrate mounted at normal incidence. The resulting thin film thickness is measured on stylus surface profiler to be 475 nm. During EDX measurements the voltage is reduced to 10 kV in order to concentrate the electron beam/specimen interaction within the film at two separate locations. The spot size is increased 
such that the CPS $>$ 15K while maintaining a DT $<$ 30\%. Peak identification was performed in the EDAX Genesis 2000 software which detected Ge$_3$Sb$_2$Te$_4$ where small traces of Si from the measurement was neglected.\\

\noindent \textbf{\large S2: E-beam evaporated GST at oblique incidence without substrate rotation (i.e. Oblique Angle Deposition (OAD))}

\noindent Prior to investigating the helical geometry, we first investigate the impact on tilt angle by performing a series of depositions with varied angle. By tilting the substrate during deposition such that the incident flux of atoms is oriented at a steep angle, porous nanorod thin films grow through a shadowing effect created by the incident adatoms. 
\begin{figure*}[ht!]
\centering
\includegraphics[width=0.8\linewidth]{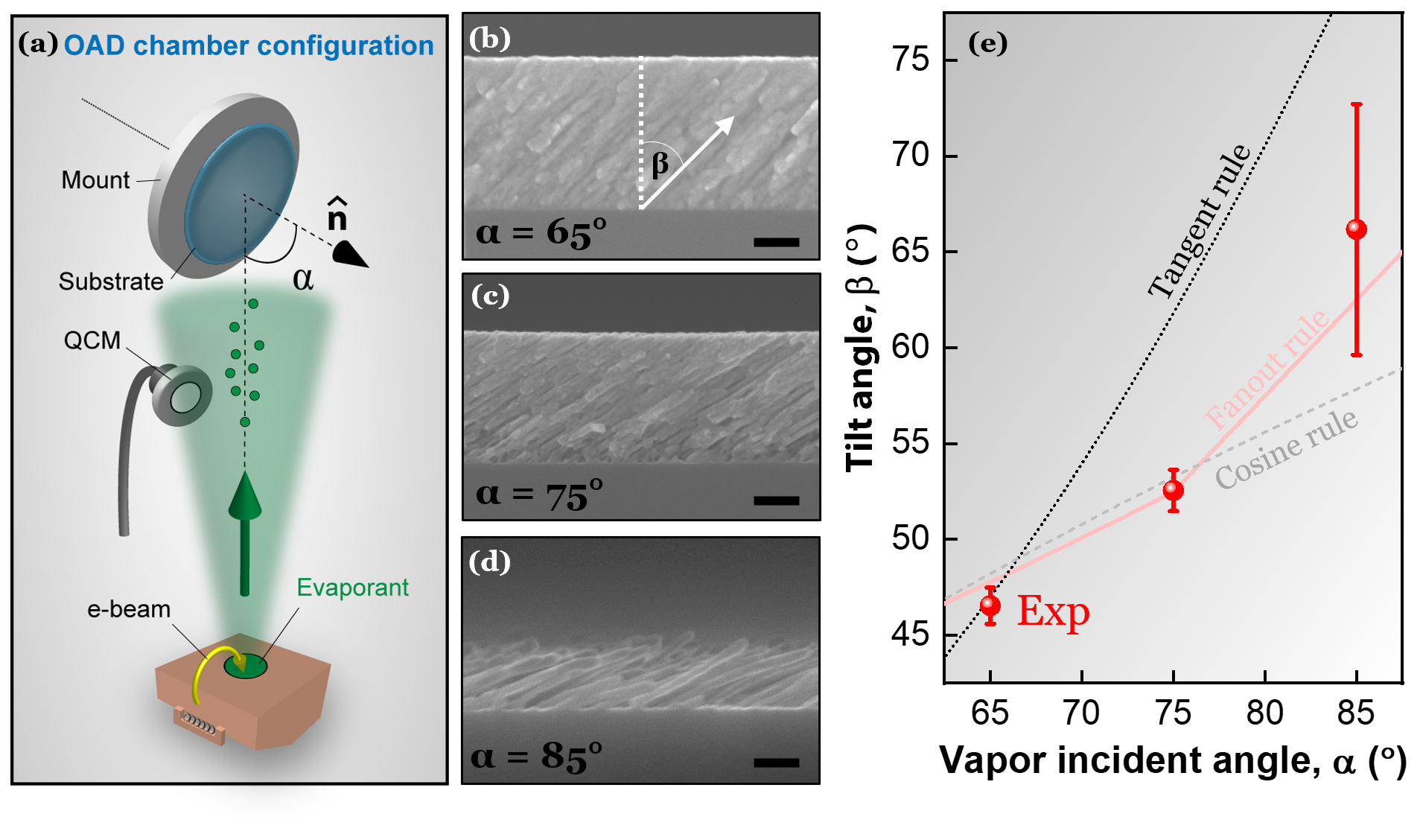}
\caption{Oblique angle deposition (a) Schematic diagrams of chamber configuration, (b)-(d) SEMs depicting the film morphology for increasing vapor incident angles, and (e) Measured collumun tilt angle as a function of $\alpha$.} 
\label{fig:conceptandesign}
\end{figure*}
It is well known that the tilt angle of the nanocolumnar structures do not equal the incident vapor flux angle. However, Nieuwenhuizen and Haanstra proposed the tangent rule
\begin{equation}
    \tan(\alpha)=2\tan(\beta)\text{,}
\end{equation}
where $\alpha$ is the vapor incident angle and $\beta$ is the column inclination angle (i.e. nanorod tilt angle). This expression was shown to be valid for $0^{\circ}< \alpha<60^{\circ}$. Later Hodgkinson et al. improved this empirical model by adding a fitting parameter based on the material under investigation. For large oblique angles he found that another empirical formula known as the cosine rule was defined as:

\begin{equation}
    2\sin(\alpha-\beta)=1-\cos(\alpha)\text{.}
\end{equation}

It is clear in Fig 4(d) that highly distinct and porous nanostructures sufficiently form for large $\alpha$ at room temperature. Thus, we fix $\alpha = 85^{\circ}$ for engineering helical structures.\\

\noindent \textbf{\large S3: Determining the morphological structures for GLAD with continuous rotation}

\noindent We also perform a detailed study on the morphology of the cylindrical helix with respect the ratio between deposition rate and rotation rate of the substrate. Several samples were grown with varying substrate rotation speeds and deposition rates. The pitch and circumference exhibit a linear trend with respect to $\gamma$. It should be noted $\gamma$ must be greater than nanorod $d$ to avoid bifurcation which is labeled in grey as the mushrooming zone. 
\begin{figure*}[ht!]
\centering
\includegraphics[width=0.5\linewidth]{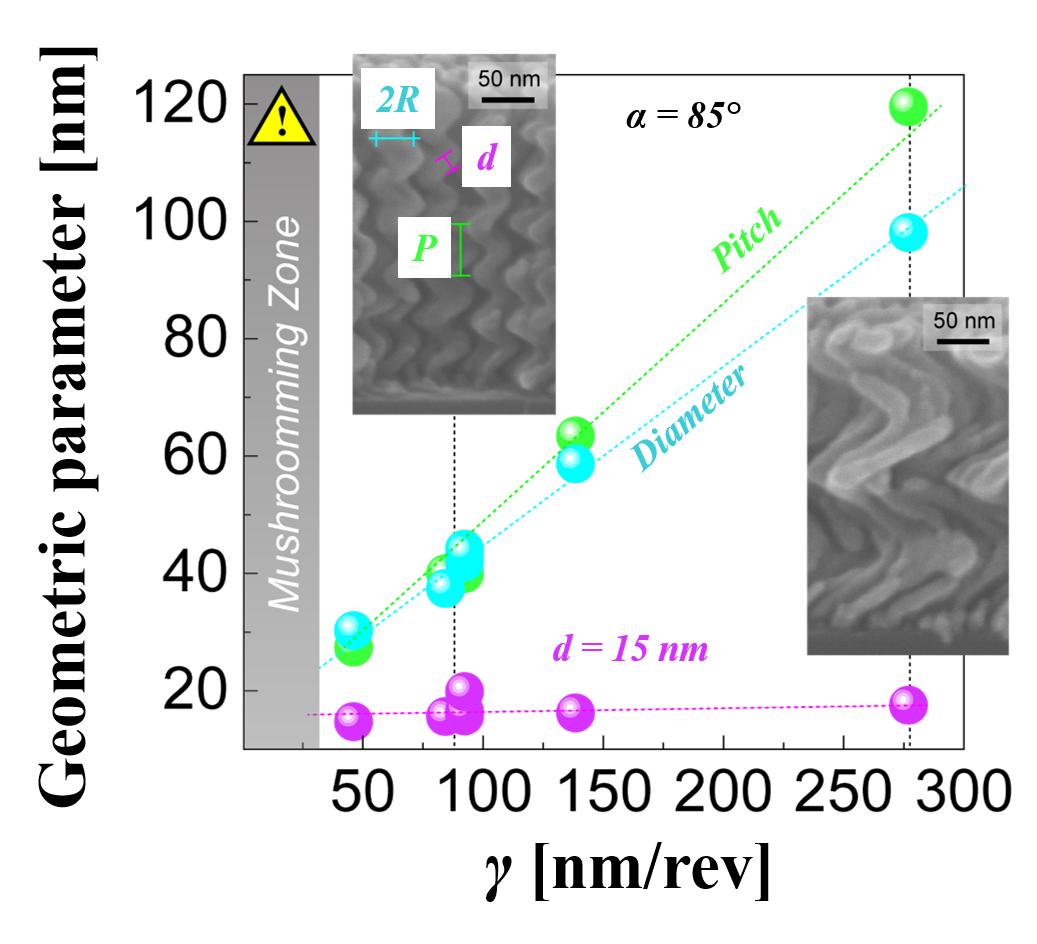}
\caption{Geometric parameters as a function of $\gamma$} 
\label{fig:conceptandesign}
\end{figure*}
The helical pitch $P$ and helical diameter $D=2R$ increase linearly with respect to $\gamma$ while the nanorod diameter, $d$ remains unaffected by varying rotation rates around 15 nm. It is worth noting that the circular dichroism results in a broader response for increasing pitching where the peak CD and CB spectral features remain at the same photon energies. \\

\noindent \textbf{\large S4: FDTD Numerical Simulations}

\begin{figure*}[ht!]
\centering
\includegraphics[width=1.0\linewidth]{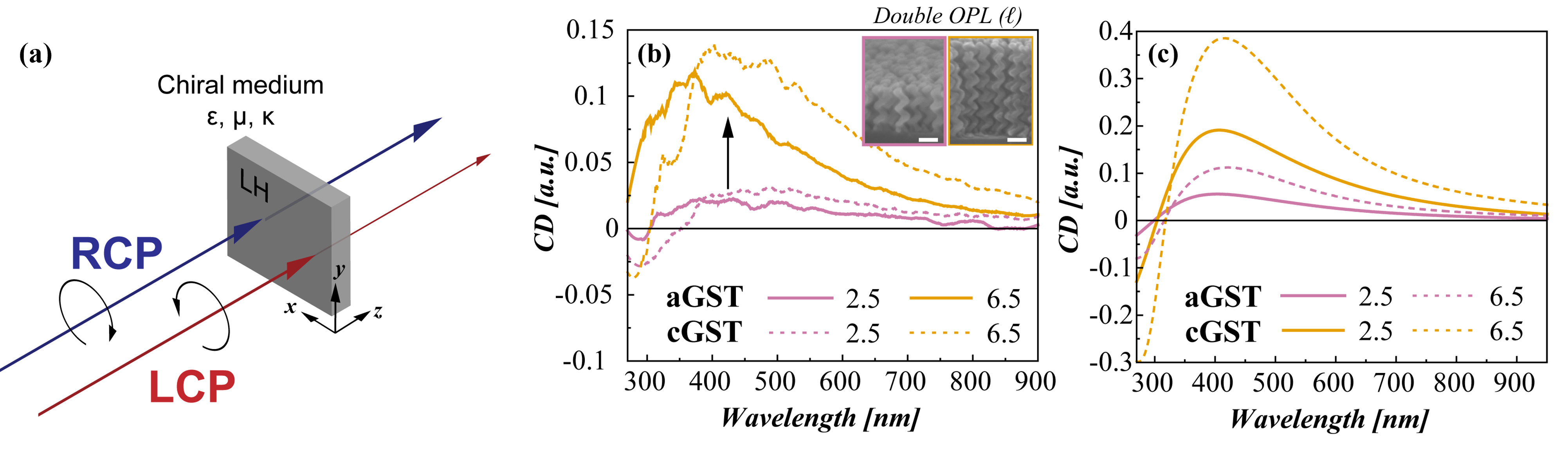}
\caption{Comparing numerical simulations to experimental spectral measurements} 
\label{fig:conceptandesign}
\end{figure*}
\noindent A full wave finite-difference time-domain (FDTD) numerical analysis (Lumerical, Inc.) is implemented to model the chiral response of the chiral PCNM medium. To ensure the accuracy of the numerical model, we first implement the optical material properties from the spectroscopic ellipsometry measurement on isotropic GST. For the numerical modal, periodic boundary conditions were imposed to model the results and while this does not perfectly reflect the aperiodic geometric, we perform a sweep on the periodic structure within the short range order of the nanostructures as well as a unit cell consisting of an array of 14 chiral absorbers. 

A full parameter sweep was conducted to access the CD signal strength as a function of number of revolutions in GST chiral nanorods with $P = 2R = 45$ nm.
Given the aperiodicity in the resulting films, we access the packing density with result to the related transmitted CD. We sweep the packing density by increasing the square periodicity ($\Lambda$) between 40 and 80 nm which spans the short range order of the nanostructures observed in top down SEM images.  

\begin{figure*}[ht!]
\centering
\includegraphics[width=1\linewidth]{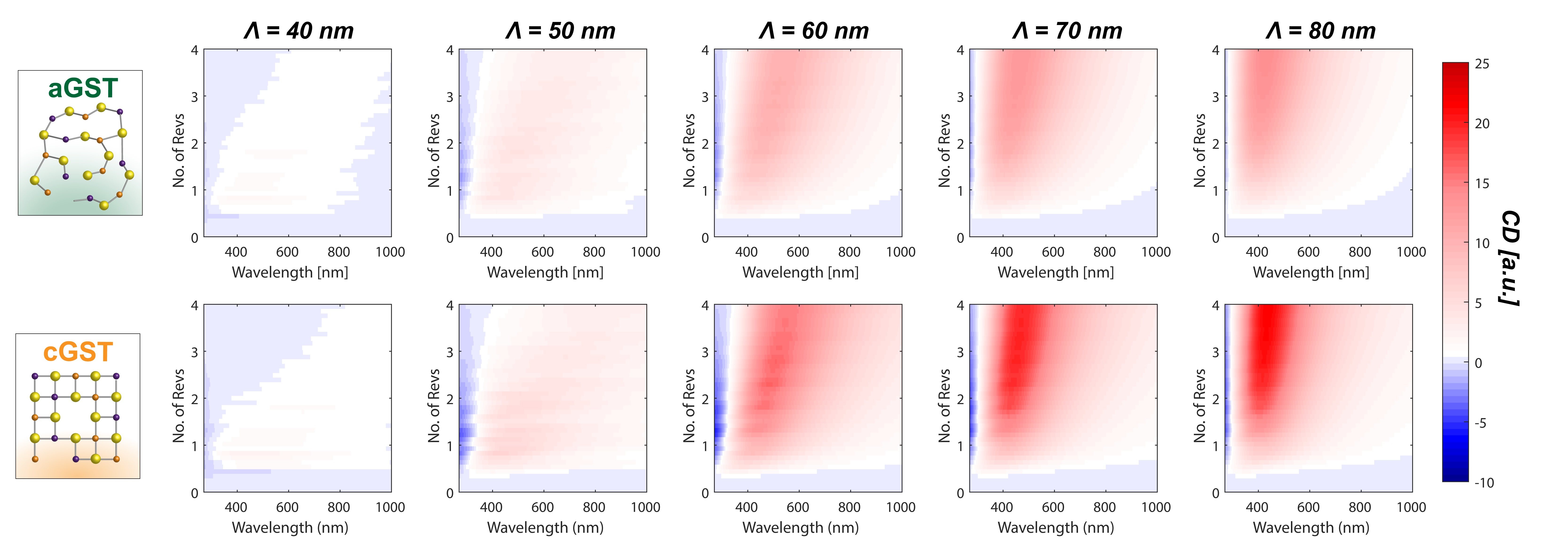}
\caption{Top and bottom bottom rows are parameter evolution plots for amorphous and crystalline states.} 
\label{fig:conceptandesign}
\end{figure*}

 We observe a monotonic evolution in the broad CD signature near 410 nm in the amorphous state (top row) which is further enhanced in the crystalline state (bottom row). For the case where the period $\Lambda$ = 40 nm, we see an attenuated CD response due to the simulation bounds not capturing the the full nanostructure with pitch P = 2R = 45 nm. Additionally, we perform a supercell simulation including 16 randomly oriented nanostructures and the spectral CD response followed the same trend.

\begin{figure*}[ht!]
\centering
\includegraphics[width=0.75\linewidth]{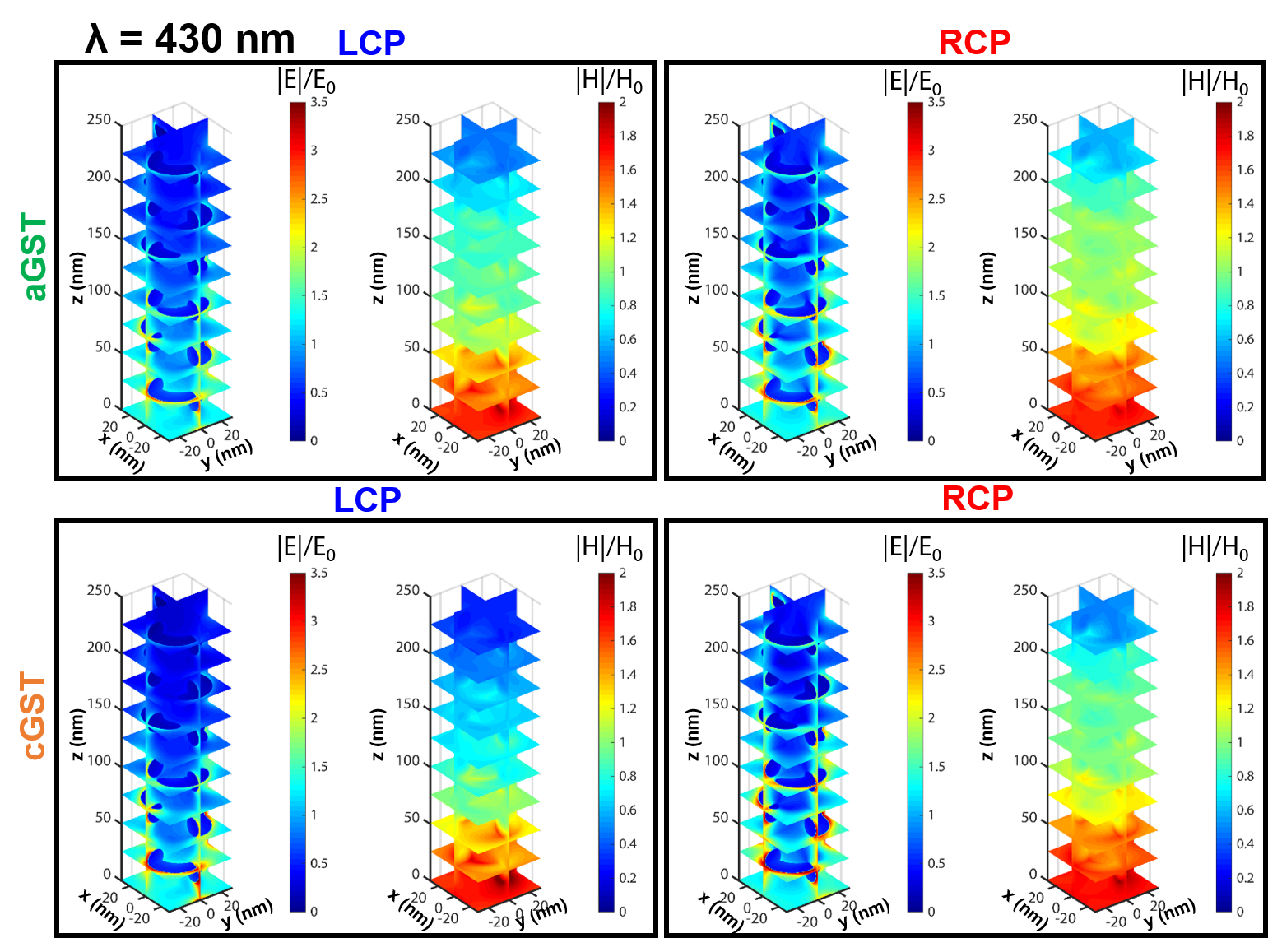}
\caption{Amorphous (top) and crystalline (bottom) E and H field distributions under LCP and RCP light excitations.} 
\label{fig:conceptandesign}
\end{figure*}

\vspace{5pt}
\hspace*{5mm}
We perform a detailed mode analysis at the peak CD to further understand the chiral absorptive phenomena. The E and H field enhancements are shown for LCP and RCP in each state where again the top and bottom rows are for aGST and cGST respectively. We observe no electrical field inside the rod but E enhancement on the surface of the nanorods that exponentially decays along the direction of propagation. For all simulations, we excited the nano-moleules from the bottom. In the H field case, we observed H enhancement in the lower region of the entire volume (inside and outside the high index structure) which exponentially decays along the propagating direction.
\vspace{5pt} 


\noindent \textbf{\large S5: Full Mueller Matrix of Chiral Structures}

\noindent We present the full Mueller Matrix of the chiral medium with the following morphological structures: $N=5$, $P=40$ nm. The nanorods with diameters of 15 nm are displaced on a double sided polished Al$_2$O$_3$ substrate. The green and gold traces are for the as-deposited amorphous state and poly-crystalline state, respectively. The results are consistent with chiral medium, in that the strongest signals are observed in the anti-diagonal elements of the Mueller Matrix. There exists some residual signal in the cGST state of the $m_{12}$. We also note there is minimal observation of depolarization at normal incidence.

\begin{figure*}[ht!]
\centering
\includegraphics[width=0.65\linewidth]{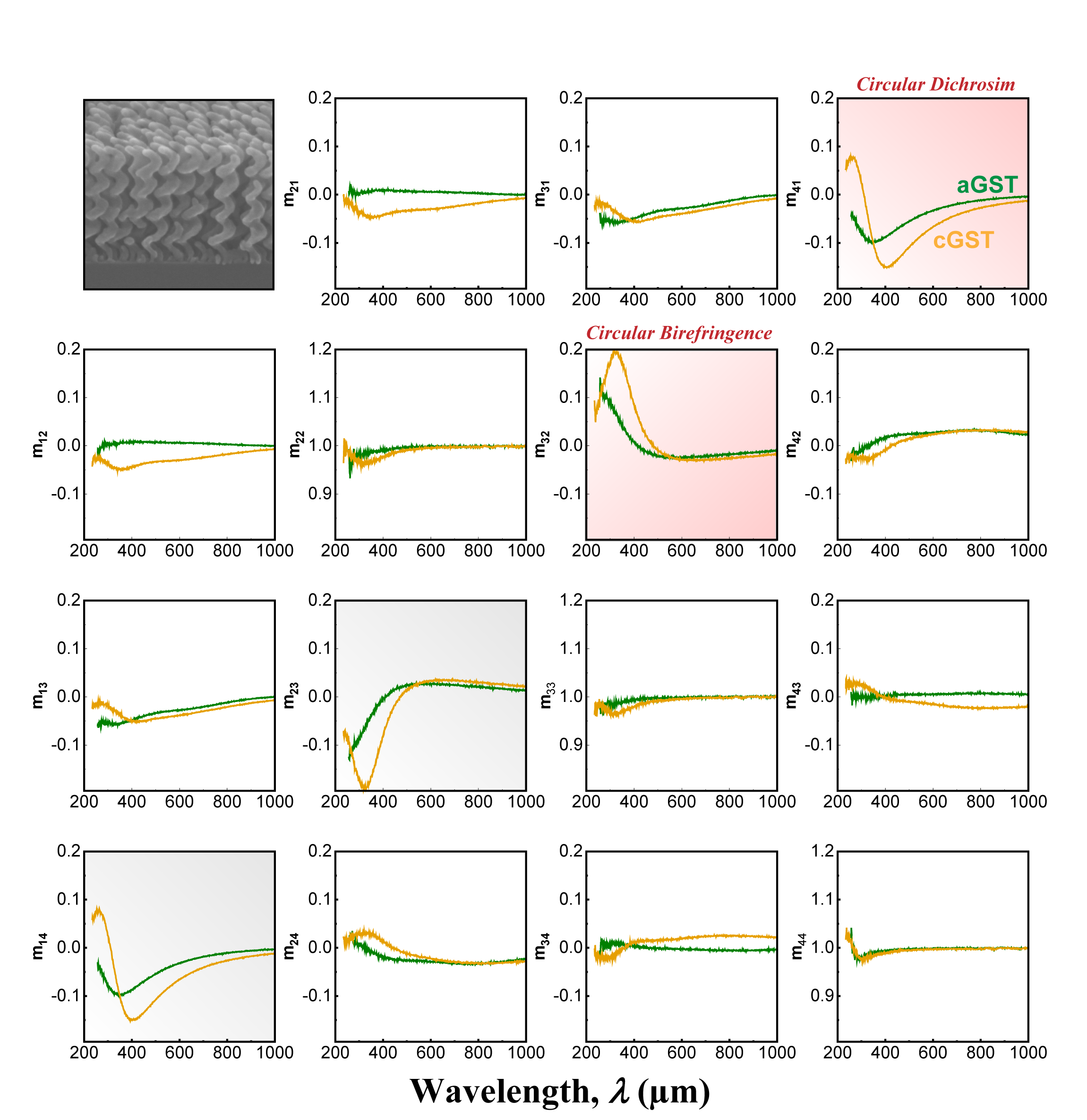}
\caption{Full transmissive Mueller Matrix ellipsometric measurements on N=5, P=40nm, RH nanorods} 
\label{fig:MM}
\end{figure*}
\newpage
\noindent \textbf{\large S6: Lifetime Measurements}

\noindent 

For lifetime measurements a pulse programming scheme consisting of 4 crystallization pulses of 15 $\mu$s duration followed by a single high fluence 300 ns amorphousation pulse was implemented. The AOM programmed 532 nm pump laser was defocues by 3 $\mu$m for the measurements where the transmitted coaligned 633 nm was focused onto a photodetector. Each oscilloscope screen capture displays 25,000 cycles each where the yellow trace is the electrical input to the AOM and blue is the transmitted 633 nm probe measured on Si detector.

\begin{figure*}[ht!]
\centering
\includegraphics[width=0.9\linewidth]{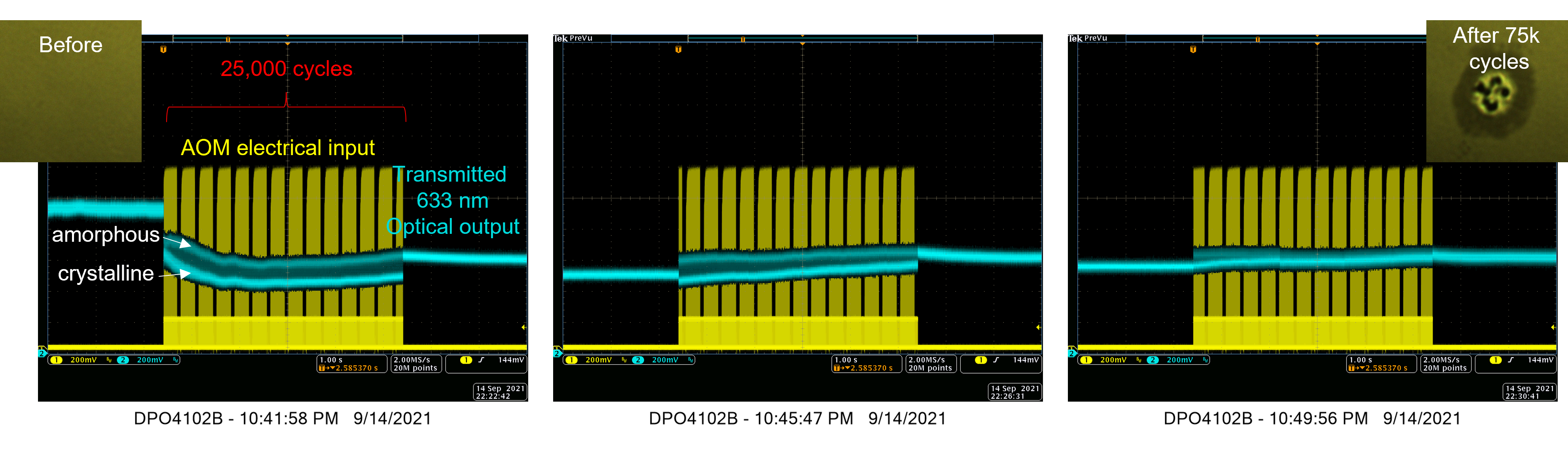}
\caption{Three screen captures of lifetime measurements.} 
\label{fig:MM}
\end{figure*}

\newpage
\noindent \textbf{\large S7: Electro-thermal Switching}

By employing the same laser used for opto-thermal switching at higher powers, laser-induced ablation of the ITO capping layer enables us to create nanometer-scale structures based on the PCNM platform in a maskless direct laser writing fashion. In particular, we create a microstrip device in which the ITO layer on top of the PCNM layer acts as a microheater, enabling the demonstration of a $7~\mu $m $\times~7~\mu$m electrically actuated pixel with the PCNM layer underneath switched between its amorphous and crystalline states using  $25~\mu$s pulse as shown in Fig. 9.

\noindent 

\begin{figure*}[ht!]
\centering
\includegraphics[width=0.9\linewidth]{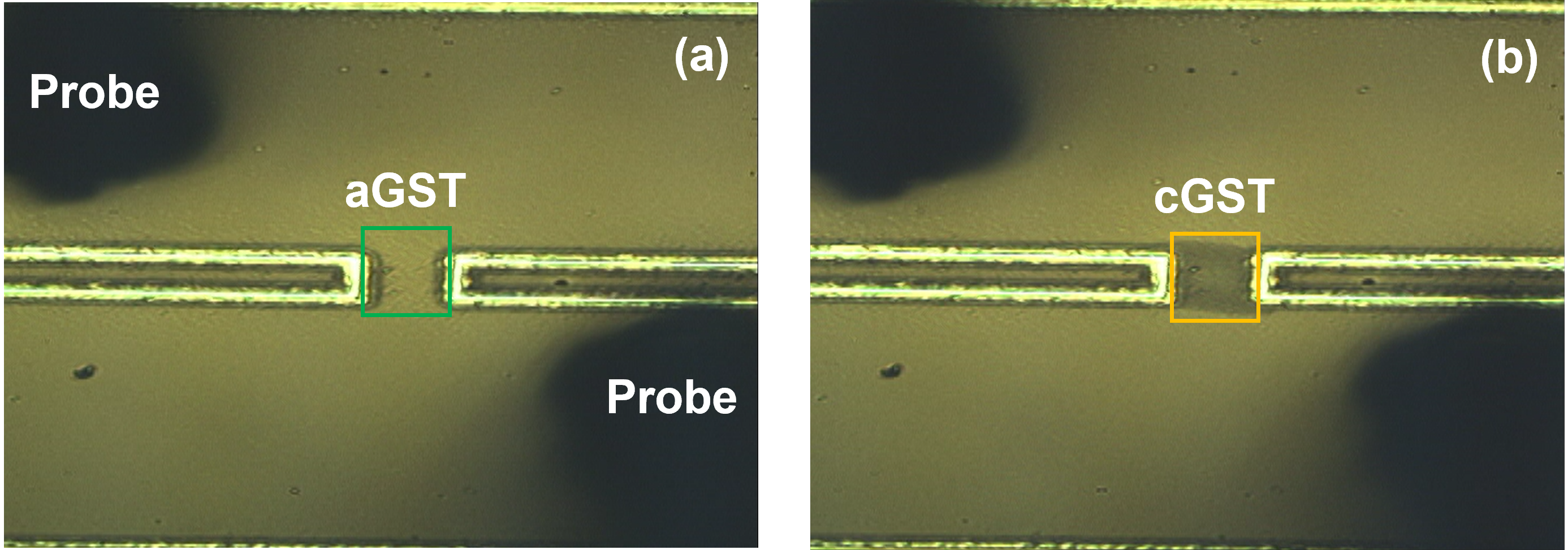}
\caption{(a) aGST and (c) cGST in maskless patterned microstrip device} 
\label{fig:MM}
\end{figure*}

\end{document}